\documentclass[sigconf]{acmart}
 
\usepackage{amsmath,amssymb,amsfonts}
\usepackage{graphicx}
\usepackage{textcomp}
\usepackage{xcolor}
\usepackage{hyperref}
\usepackage{cleveref}
\usepackage{longtable}      
\usepackage{array}          
\usepackage{booktabs}       
\usepackage{lscape}         
\usepackage{algorithm}     
\newcommand{\EndIndent}{}  
\usepackage[noend]{algpseudocode} 
\usepackage{listings}
\usepackage{etoolbox}
\newcommand{\Indent}{\hspace{2em}} 
\makeatletter
\preto{\@verbatim}{\small}
\makeatother
\usepackage{totpages}
\usepackage{framed}
\definecolor{diffremoved}{RGB}{255,221,221}  
\definecolor{diffadded}{RGB}{221,244,221}    
\definecolor{diffremovedtext}{RGB}{0,0,0}  
\definecolor{diffaddedtext}{RGB}{0,0,0}    

\newcommand{\codesmall}[1]{\texttt{\footnotesize #1}}

\newcommand{\diffremoved}[1]{%
\vspace{0.5ex}
\noindent\colorbox{diffremoved}{\color{diffremovedtext}- \codesmall{#1}}%
}

\newcommand{\diffadded}[1]{%
\noindent\colorbox{diffadded}{\color{diffaddedtext}+ \codesmall{#1}}%
\vspace{0.5ex}
}

\newcommand{\beforeafter}[2]{%
\diffremoved{#1}\\%
\diffadded{#2}%
}

\newcommand{\beforeaftermultiline}[2]{%
\vspace{0.5ex}
\noindent\begin{minipage}{\linewidth}%
\colorbox{diffremoved}{\parbox{0.98\linewidth}{\color{diffremovedtext}\footnotesize\ttfamily\raggedright- #1}}\\[2pt]%
\colorbox{diffadded}{\parbox{0.98\linewidth}{\color{diffaddedtext}\footnotesize\ttfamily\raggedright+ #2}}%
\end{minipage}%
\vspace{0.5ex}
}

\newcommand{\PatternItem}[2]{%
  \vspace{1ex}\noindent\textbf{#1} #2 %
}
\newcommand{\eg}{\textit{e}.\textit{g}., }

\newcommand{\etc}{\textit{etc}.}

\copyrightyear{2026}
\acmYear{2026}
\setcopyright{cc}
\setcctype{by}
\acmConference[MSR '26]{23rd International Conference on Mining Software Repositories}{April 13--14, 2026}{Rio de Janeiro, Brazil}
\acmBooktitle{23rd International Conference on Mining Software Repositories (MSR '26), April 13--14, 2026, Rio de Janeiro, Brazil}
\acmPrice{}
\acmDOI{10.1145/3793302.3793334}
\acmISBN{979-8-4007-2474-9/2026/04}

\title{Source Code Hotspots:\\
A Diagnostic Method for Quality Issues}

\author{Saleha Muzammil}
\orcid{0009-0009-6097-592X}
\affiliation{%
  \institution{University of Virginia}
  \city{Charlottesville}
  \country{USA}
}
\email{saleha@email.virginia.edu}

\author{Mughees Ur Rehman}
\affiliation{%
  \institution{Virginia Tech}
  \city{Blacksburg}
  \country{USA}
}
\email{mughees@vt.edu}

\author{Zoe Kotti}
\orcid{0000-0003-3816-9162}
\affiliation{%
  \institution{Athens University of Economics and Business}
  \city{Athens}
  \country{Greece}
}
\email{zoekotti@aueb.gr}

\author{Diomidis Spinellis}
\orcid{0000-0003-4231-1897}
\affiliation{%
 \institution{Athens University of Economics and Business}
  \city{Athens}
  \country{Greece}
}
\email{dds@aueb.gr}
\affiliation{%
  \institution{Delft University of Technology}
  \city{Delft}
  \country{Netherlands}
}

\begin{document}

\begin{abstract}
Software source code often harbours ``hotspots''---small portions of the code
that change far more often than the rest of the project
and thus concentrate maintenance activity.
We mine the complete version histories of 91 evolving, actively developed GitHub repositories
and identify 15 recurring line-level hotspot patterns that explain why these hotspots emerge.
The three most prevalent patterns are \emph{Pinned Version Bump} (26\%), revealing brittle release practices;
\emph{Long Line Change} (17\%), signalling deficient layout; and
\emph{Formatting Ping-Pong} (9\%), indicating missing or inconsistent style automation.
Surprisingly, automated accounts generate 74\% of all hotspot edits,
suggesting that bot activity is a dominant---but largely avoidable---source of noise in change histories.
By mapping each pattern to concrete refactoring guidelines and continuous integration checks,
our taxonomy equips practitioners with actionable steps
to curb hotspots and systematically improve software quality in terms
of configurability, stability, and changeability.
\end{abstract}
\maketitle


\section{Introduction}
A key advantage of software over hardware is
that it is considerably more malleable.
Yet software changes come at a cost:
they require effort to develop, test, and deploy them,
and they may introduce faults or lead to failures.
Consequently, it is profitable to improve
our understanding of why software is changing and
our management of software changes.

Inspired by an electrical engineering diagnostic method where
infrared thermography is employed to assess equipment reliability~\cite{JST12}
and identify defects~\cite{HMT13},
we propose and study source code hotspots as a way to identify
software engineering defects.
We term frequent localized changes to specific software source code
lines as \emph{source code hotspots} (hotspots from now on).

We posit that when a hotspot appears,
it often signals a deeper pathology in the codebase. The root cause can be a weak software architecture, a flawed design, inadequate tooling, or a deficient
software development process.
Frequently, the underlying issue fits an anti-pattern~\cite{brown1998},
such as Continuous Obsolescence~\cite[p. 85]{brown1998}
or Stovepipe Enterprise~\cite[p. 147]{brown1998}.
For example,
a hard-coded currency exchange rate will require frequent updates to its value.
A better design would
have that value stored in a database or configuration file
or retrieve it through a web service.

From a theoretical perspective, our research focuses on software quality,
particularly on maintainability.
Maintainability is a critical attribute
encompassing, among others, three interrelated dimensions:
\emph{configurability}, \emph{stability}, and \emph{changeability}~\cite{Spi06}.
\emph{Configurability} denotes the ease with which behaviour can be adjusted via settings or external files, reducing the need for source edits. Lack of configurability forces code modifications for every slight adjustment, inflating code churn.
\emph{Stability} measures a system's resilience to unintended impacts when changes occur elsewhere; unstable systems incur cascading edits across modules. Finally, \emph{changeability} captures how simply modifications (\eg feature additions or bug fixes) can be implemented; sub par changeability leads to repetitive corrections of the same lines.
Software code that changes often may
lack in changeability,
miss configurability options, or
exhibit instability when modified.
Deficiencies in any dimension can manifest as hotspots.

An additional challenge is that modern software evolution is increasingly shaped by automation.
 Bots routinely perform dependency updates, metadata regeneration, and formatting changes, inflating code churn 
 and potentially obscuring developer-driven maintenance effort. In our context, bot activity refers to commits authored 
 by automated accounts recorded in Git history. Without distinguishing such 
 automated changes from human-authored edits, hotspot analyses risk conflating mechanical noise with meaningful software evolution, 
 making it difficult to identify which patterns reflect genuine maintainability issues versus routine mechanical updates.


Unlike previous research that primarily focused on detecting anti-patterns
arising from design principle violations at coarse granularity (e.g., files or commits)~\cite{JAJ12,PDBO14,MCKX21},
our work pioneers the identification of \emph{line-level} code churn.
We developed and employed an algorithm and a corresponding tool
that perform fine-grained analysis
of revision histories to track the evolution of individual lines over time.
This line-level granularity isolates the exact loci of repeated change,
enables actionable diagnosis of distinct churn sources
(e.g., configuration constants versus formatting noise), and allows attribution of hotspot edits 
to human developers versus automated accounts based on committer identity.
By zooming in on these frequently altered lines, we can not only relate them to
design or process anti-patterns but also pinpoint , avoidable changes.
Thus, our identification of hotspots offers a concrete strategy for improving
software quality.
To achieve these goals, we have formulated the following research questions.

\begin{description}
    \item[RQ1:] Which technical or process factors most often give rise to
    hotspots, and what concrete practices can curb their recurrence?
    \item[RQ2:] How do hotspots vary in scope, persistence, and location (\eg administrative files vs.\ executable code) across diverse projects?
    \item[RQ3:] How do bot-authored commits influence hotspots, and what specific anti-patterns are predominantly bot-driven?

\end{description}

This study contributes the following:
a novel method for identifying a class of software and process defects
based on (line-level) hotspots,
a tool for identifying hotspots through the granular analysis and
propagation of changes in revision histories;
a taxonomy of reasons that give rise to hotspots;
metrics regarding the occurrence of hotspots in a stratified sample of
open-source software projects; and,
an examination of the role bots play in the creation of hotspots.

The remainder of this paper is organized as follows.
Section~\ref{sec:related} reviews prior work on code churn, change taxonomies, and the role of automation in software evolution.
Section~\ref{sec:method} describes our methods, including repository sampling, hotspot detection,  taxonomy construction, and bot identification.
Section~\ref{sec:results} presents our hotspot taxonomy, quantitative characteristics of hotspot occurrence, and comparative analysis of bot-generated versus human-authored changes.
Section~\ref{sec:discussion} discusses the implications of our findings for software maintainability and development practice.
Finally, Sections~\ref{sec:threats} and~\ref{sec:conclusion} outline threats to validity and conclude.

\section{Related Work}
\label{sec:related}
We organize our review around four key research threads
that inform our contributions:
(1) the foundational understanding of code churn and source code hotspots
in predicting software quality;
(2) existing taxonomies of software smells and anti-patterns
that motivate our line-level approach;
(3) technical advances in mining code churn; and
(4) the emerging influence of automated bots on software development practices.

\subsection{Code Churn and Fault Prediction}
The relationship between code churn and software quality has been extensively studied, establishing churn as a reliable predictor of fault-prone areas. Early work by \citet{graves2002predicting} demonstrated that files with high modification frequencies correlate strongly with defect density, a finding corroborated by \citet{nagappan2005use} and \citet{kim2007predicting} across multiple industrial projects.
\citet{hassan2009} advanced this line of research by introducing temporal complexity metrics, showing that the \emph{pattern} of changes matters as much as the volume.
\citet{giger2012} analyzed types of code changes, further illustrating that specific churn types can predict future modifications.
Beyond defect-focused analyses, the shape of project activity matters: the distribution of commit frequency in open source projects is highly skewed, with a small fraction of contributors and periods accounting for a disproportionate share of commits~\citep{KRS13}. This heavy-tailed activity further motivates our investigation of line-level hotspots.
Empirical studies have characterized \emph{what} developers change and \emph{why}. For example, work on change-reason analysis uses historical repositories to infer the intent behind modifications~\cite{mockus2000identifying}. Other studies propose and analyze taxonomies of change types and patterns~\cite{fluri2008discovering,islam2021changes,trautsch2023really,jaafar2017analyzing}, including mining unknown change patterns from code histories~\cite{negara2014}. 

However, these studies predominantly are defined
at file or module granularity, leaving a gap in understanding the specific \emph{lines} that drive hotspot behavior. Our work addresses this limitation by providing the first systematic analysis of line-level code churn patterns across a stratified sample of repositories and the idea of using hotspots as a diagnostic method for software quality issues.
\subsection{Software Smells, Anti-Patterns, and Change Taxonomies}
Code smells~\cite{fowler2018} and design anti-patterns
\cite{brown1998,webster1995,koenig1998}
signal maintainability debt. At code level, Fowler's seminal catalogue (\emph{God Class}, \emph{Long Method}, \emph{Feature Envy}) formalised
``smells'' as maintainability warning signs~\cite{fowler2018}. This catalogue has since expanded to encompass architectural anti-patterns~\cite{brown1998,webster1995,koenig1998}, with \citet{mo2019} identifying structural problems like \emph{Unstable Interface} and \emph{Modularity Violation}.
In this area,
\citet{garcia2009} contributed systematic detection approaches
for architectural smells,
\citet{SAKS20} identified co-changes as a way to detect architectural smells,
and
\citet{MCKX15} proposed explicitly the employment of hotspot patterns
to identify architecture smells.

Modern software development has introduced new categories of smells.
\citet{taibi2018} catalog microservice-specific anti-patterns,
\citet{sharma2016does} identify configuration smells, showing that configuration smells exhibit higher co-occurrence than implementation smells,
while \citet{PHBE22}, in common with our goals,
describe the detection of process anti-patterns from project data.

Parallel to structural smell research, change-focused taxonomies have emerged. Early frameworks by \citet{mens2003towards} and \citet{buckley2005towards} established theoretical foundations for classifying software evolution patterns. More recent work has provided empirical taxonomies~\cite{lehnert2012taxonomy, islam2021changes, palomba2017exploratory}.

Importantly for our work, \citet{nguyen2013study} demonstrated that small code changes exhibit high repetitiveness (70--100\%), with patterns remaining consistent both within and across projects. This finding directly motivates our line-level analysis, as repetitive change patterns suggest systematic maintainability issues that can be addressed through targeted interventions. 
Most existing smell and change taxonomies operate at coarse granularity
(e.g., classes, files, components, or services)~\cite{mens2003towards,buckley2005towards,lehnert2012taxonomy,islam2021changes,palomba2017exploratory}.
While useful for describing broad maintenance trends, such abstractions can conflate heterogeneous edits within the same artifact,
limiting actionability.
In contrast, our contribution provides a line-level taxonomy,
which distinguishes pathological churn from semantically meaningful edits
and thereby enables direct CI enforcement and
the proposal of concrete refactoring guidelines.

\subsection{Techniques and Tools for Detecting Code Churn}
Textual difference algorithms have traditionally been employed to capture software changes at various granularities~\cite{miller1985file, myers1986nd}. Modern approaches~\cite{falleri2014fine, falleri2024fine} offer significantly improved accuracy and granularity, facilitating the identification of specific line-level changes and thus enabling hotspot detection~\cite{asaduzzaman2013lhdiff}.

Tools built on these methods like \texttt{ChangeDistiller}~\cite{fluri2007change}, utilize tree-differencing algorithms to extract source code changes efficiently. Similarly, PYCT~\cite{lin2016empirical} extends change detection capabilities to Python codebases, reducing manual overhead in classification tasks.

While various tools exist for architectural and design anti-pattern detection (\eg SonarQube, Designite~\cite{sharma2016designite}, DV8~\cite{cai2019dv8}), there remains a notable absence of tools explicitly designed for identifying line-level code churn patterns (hotspots) from revision histories. Addressing this gap is central to our research.

\subsection{Automation and Bots}
Large-scale profiling of GitHub activity shows substantial heterogeneity in who contributes, what changes are made, and when/where activity concentrates across ecosystems,
with an increasing involvement of bots~\citep{XWWZ22}.
Building on this macro view, we focus on the automation slice of that activity, examining how bot-generated changes interact with churn and hotspots.
Automation now generates a non-trivial share of source code commits. Bots automate a range of tasks, often leading to superficial or repetitive changes that inflate churn metrics and potentially obscure substantial development efforts~\cite{erlenhov2020empirical}.

\citet{wessel2018} explored code-review bots, demonstrating that while bots streamline workflows by merging pull requests more rapidly, they simultaneously reduce developer engagement and collaborative interactions, potentially hindering meaningful software evolution.
Similarly, \citet{dey2020exploratory} found that bot-generated minor changes frequently overwhelm change histories, masking important developer-driven modifications and inflating overall churn.

Furthermore, \citet{kinsman2021software} investigated GitHub Actions, identifying an increase in pull request rejections and a reduction in merged commits post-adoption, further illustrating bots' complex influence on software development practices.
\citet{golzadeh2021ground} developed methods to systematically identify bots through behavioral analysis, reinforcing the importance of distinguishing automated activities from human-driven contributions.

In summary, prior work has (i) measured churn and change patterns largely at file, module, or commit granularity~\cite{graves2002predicting,hassan2009,giger2012}, (ii) catalogued smells and anti-patterns and, in some cases, connected them to evolution signals using change-history information~\cite{palomba2013,palomba2014mining}, and (iii) analyzed bot activity and its effects on development traces~\cite{wessel2018,dey2020exploratory,erlenhov2020empirical}. Our study complements this literature by systematically identifying recurring \emph{line-level} hotspot patterns, mapping them to a taxonomy with actionable mitigations,
and quantifying the automation share within each pattern.
\section{Methodology}
\label{sec:method}
Our study aims to identify in open-source software repositories
line-level code churn (source code hotspots):
files and corresponding lines that change much more frequently than others.
To optimize the identification of hotspots
we track modifications first at the file level and then at the line level.

\subsection{Repository Selection and Mining}

Given the more than 420 million repositories hosted on GitHub as of January 2023~\cite{dohmke2023100}, analyzing them all was clearly infeasible. 
We therefore selected projects using stratified sampling~\cite{Ney92}
based on GitHub stars as a proxy for community interest~\cite{borges2018s} and GitHub forks as a proxy for direct developer involvement to strike a balance between breadth and tractability. Without stratification, a purely random draw would be dominated by dormant or toy projects~\cite{kalliamvakou2014promises}, whereas focusing only on the most-starred repositories would conflate hotspot phenomena with the engineering cultures of a small elite. Guided by best-practice recommendations for repository curation and mining~\cite{munaiah2017curating,kalliamvakou2014promises,dabic2021sampling,vidoni2022systematic}
and building on a specific method employed by \citet{SLMS24},
we divided the population into five strata (11--100, 101--1000, 1001--10\,000, 10\,001--100\,000, and 100\,001--1\,000\,000 stars/forks) and then applied the following inclusion criteria for selecting the repositories.

\begin{itemize}
\item{More than ten GitHub stars or forks, to select projects relevant to the software engineering community, and avoid personal projects and student exercises. These two metrics are considered to be the most useful GitHub project popularity metrics~\cite{borges2018s}.}
\item{At least one commit on every half-year interval, in order to examine continuously evolving software, with changes hotspots potentially surfacing over longer periods.}
\item{At least 10\,000 commits to provide a reasonable opportunity
for hotspots to occur.}
\end{itemize} 

The metadata of the repositories was queried through the GitHub API on October 20, 2024.
To validate our inclusion and exclusion criteria, we manually inspected a random sample of ten repositories from our initial selection. This inspection confirmed that all sampled repositories matched our intended criteria regarding commit counts, creation dates, and fork status.

\begin{figure}[ht]
  \centering

  \includegraphics[width=0.49\linewidth, height=0.13\textheight]{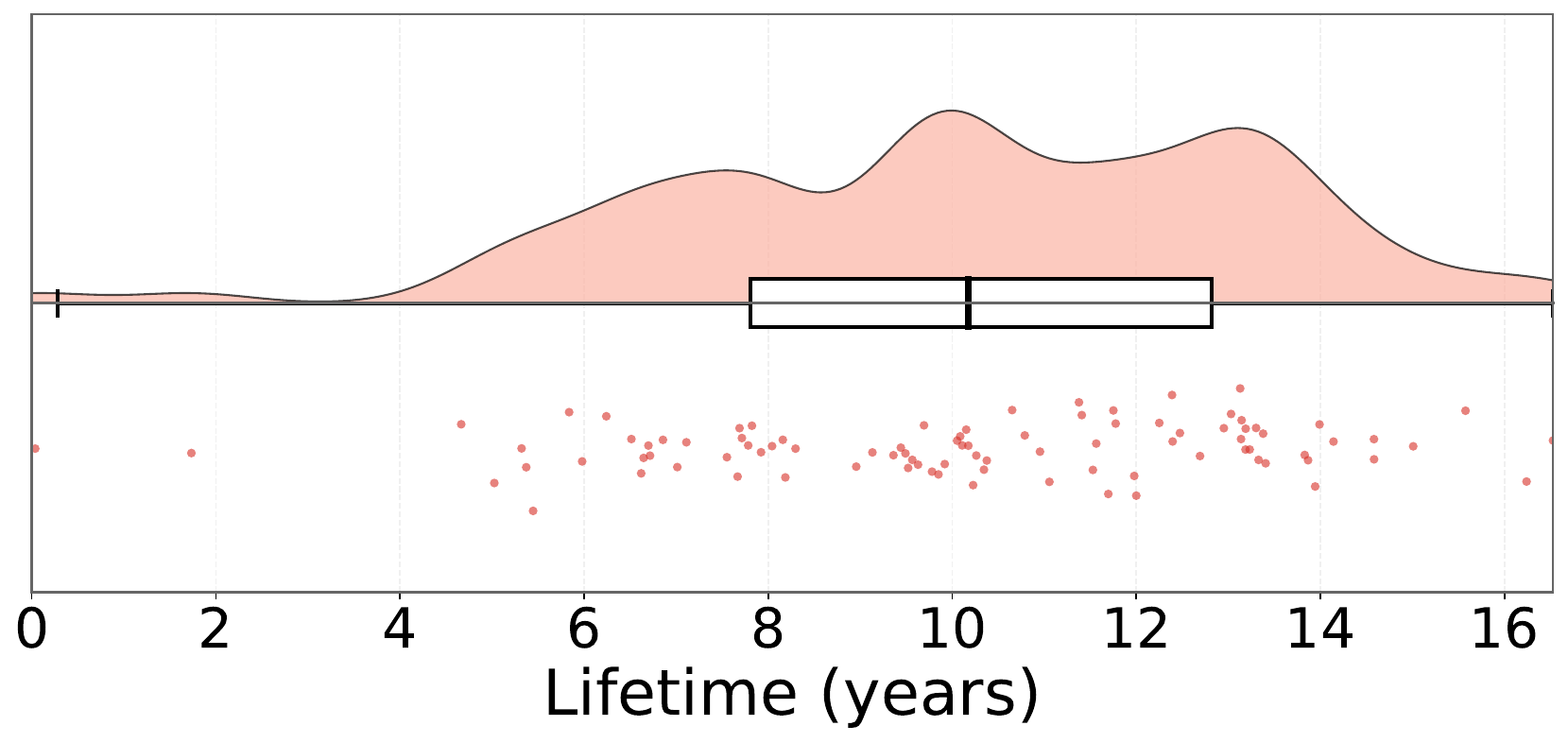}
  \includegraphics[width=0.5\linewidth, height=0.13\textheight]{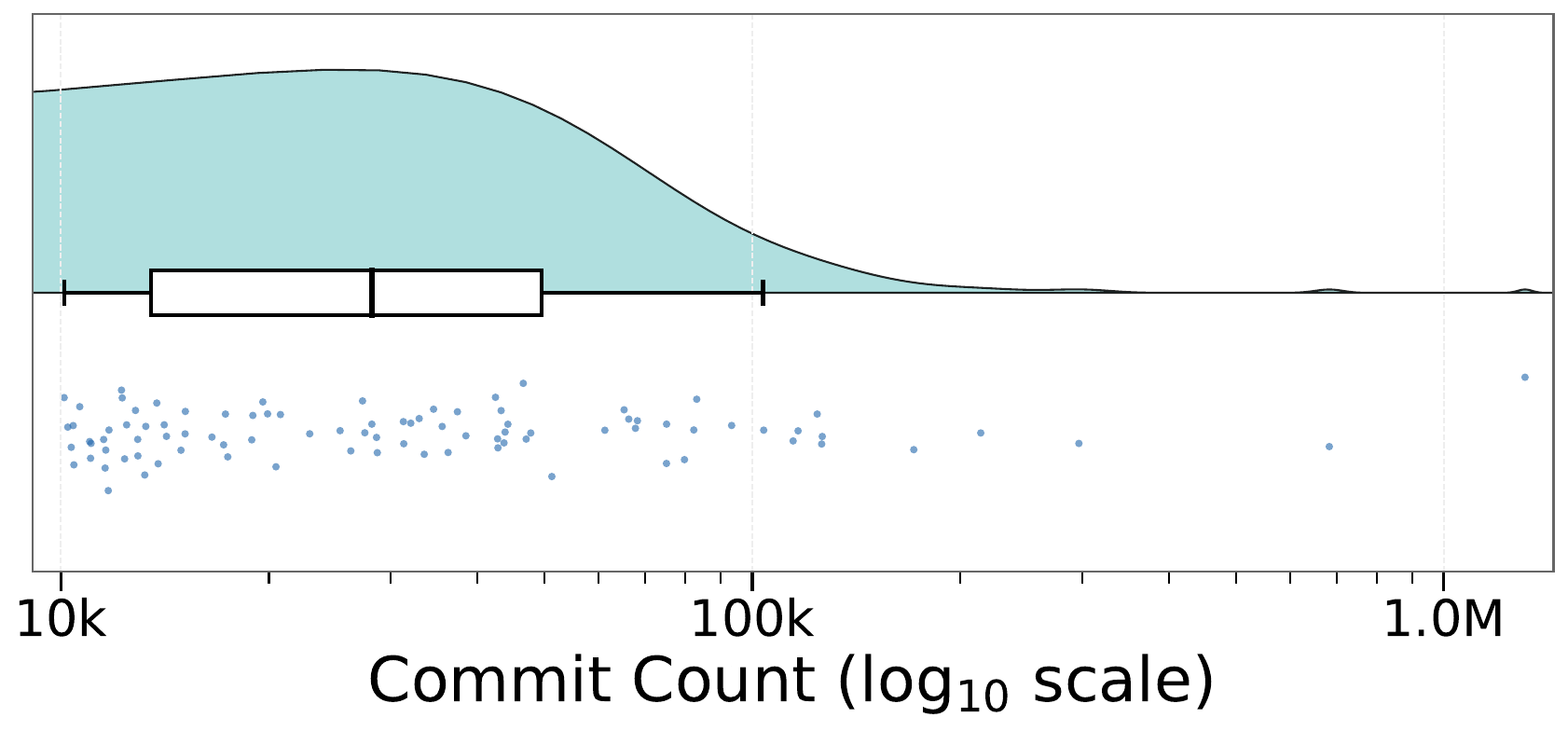}

  \vspace{1em}

  \includegraphics[width=0.495\linewidth, height=0.13\textheight]{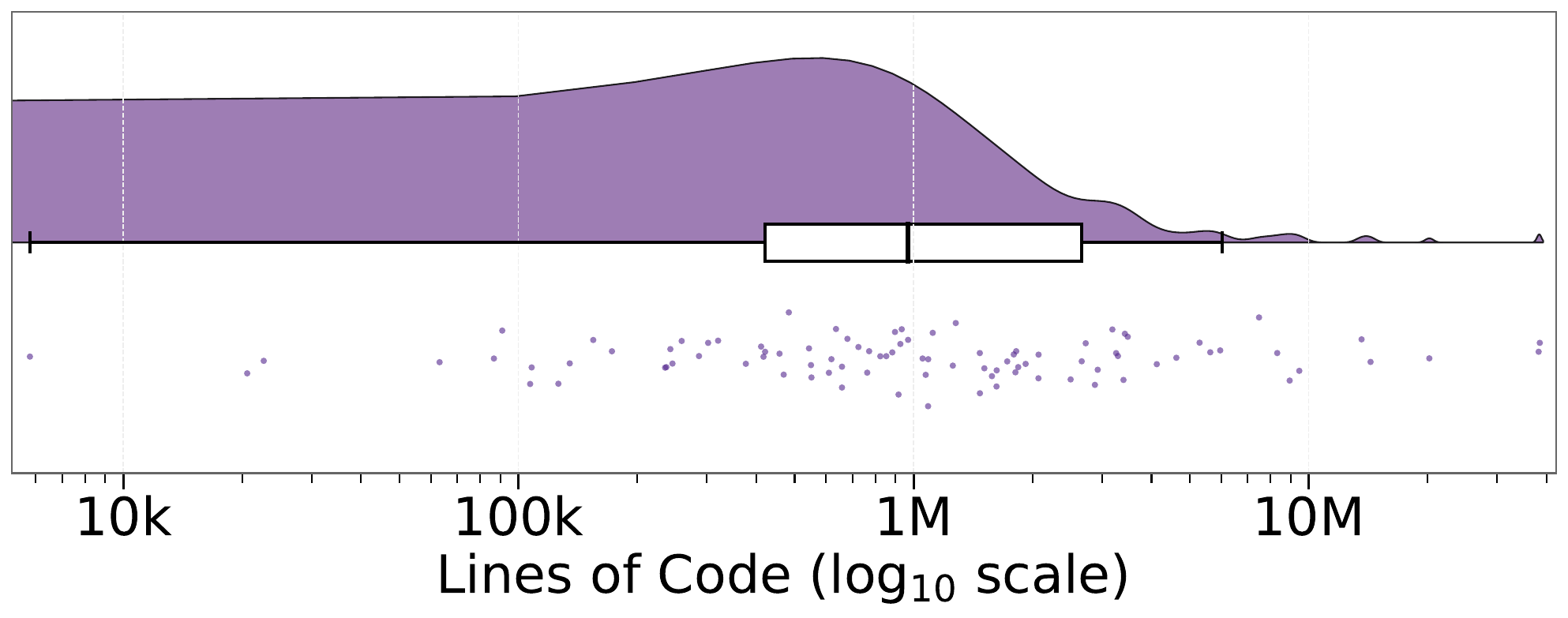}
  \includegraphics[width=0.495\linewidth, height=0.13\textheight]{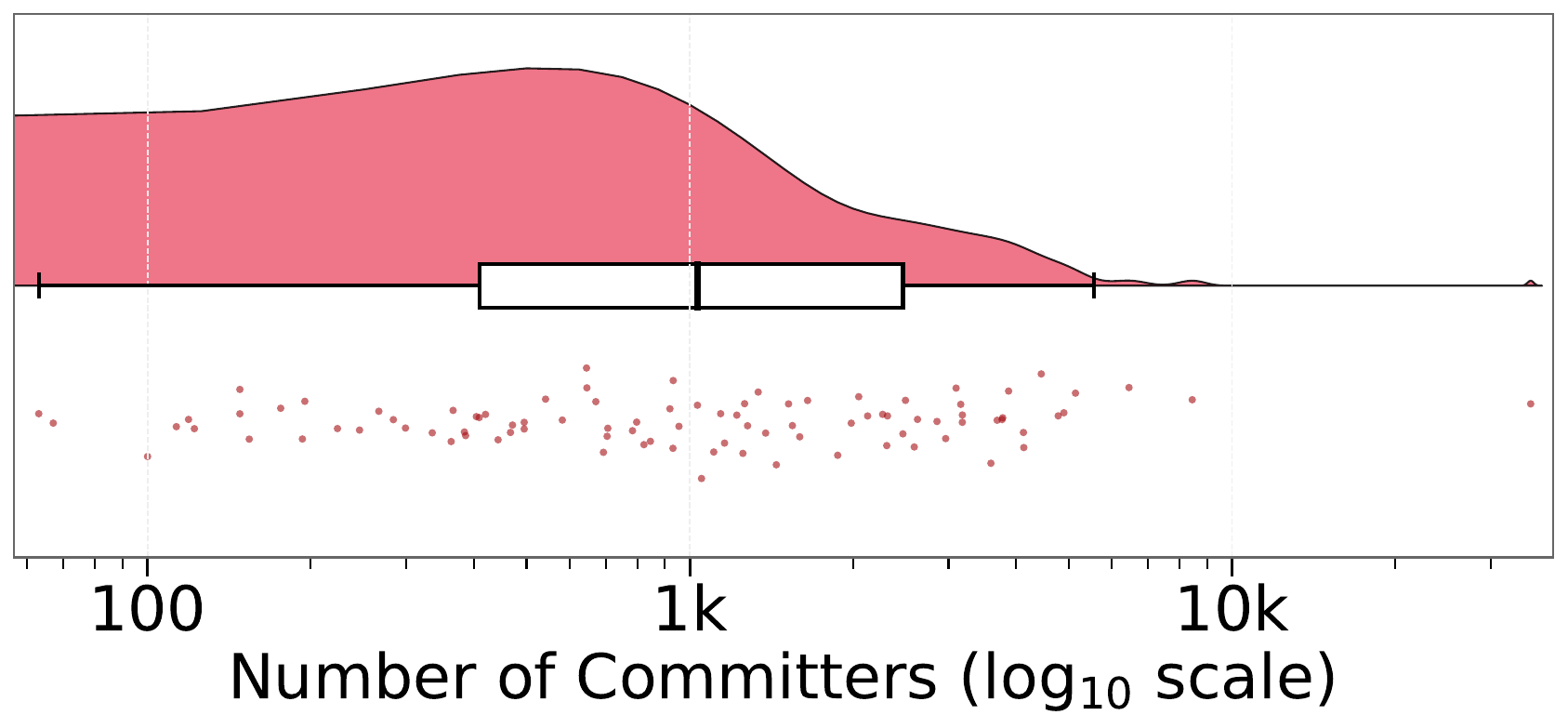}

  \caption{Lifetime (top left), commits (top right), size (bottom left), and contributors (bottom right) of analyzed repositories. We use a log$_{10}$ scale for commit count, lines of code, and contributor count to visualize heavy-tailed distributions.}
  \label{fig:ghstats}
\end{figure}
\begin{figure}[t]
  \centering
  \includegraphics[width=\linewidth]{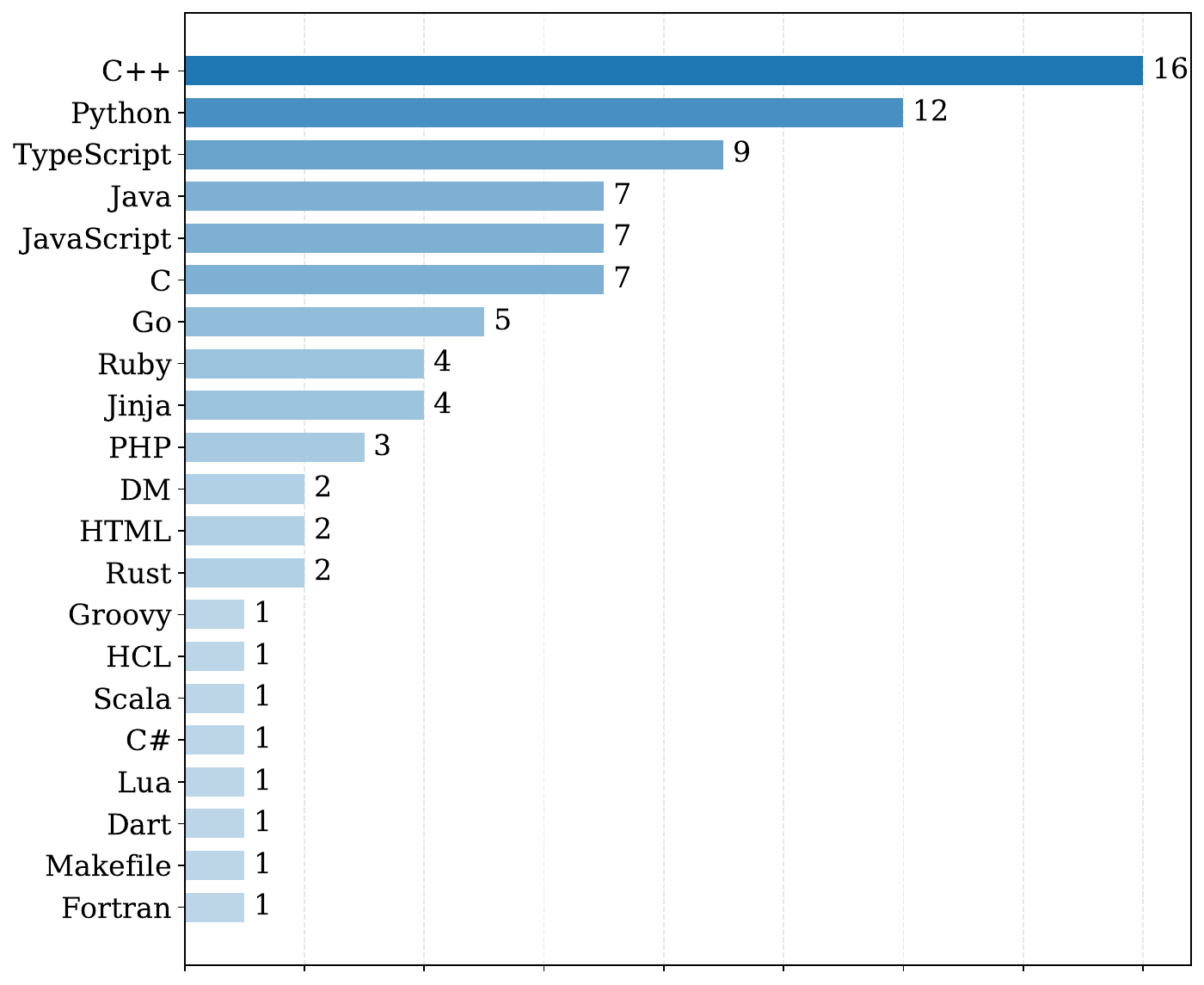}
  \caption{Language Distribution of Projects}
  \label{fig:language}
\end{figure}

Our final dataset consisted of 91 repositories, spanning diverse sizes and activity levels;
see the rainclouds~\cite{APWM21} in~\Cref{fig:ghstats}
depicting metric distributions, inter-quartile ranges, medians,
and the actual data points.
The long tails in commits, size, and committers are the expected result
of the employed stratified sampling.
Repository ages ranged from 0.04--16.53 years, with commit counts ranging from 10\,114--1\,310\,015. The number of source code lines ranged from 5\,805--38\,433\,291 lines, while contributor counts ranged from 63--35\,543.  The repositories represented a wide range of programming languages (see~\Cref{fig:language}), with C++ being the most prevalent (16 repositories), followed by Python (12 repositories) and TypeScript (9 repositories). Java, JavaScript, and C each contributed seven repositories. Other languages including Go, Ruby, specialized languages such as Jinja and PHP, and emerging systems languages like Rust were also represented, ensuring coverage across different programming paradigms and domains.
Table~\ref{tab:repo-categories} summarizes the domain distribution of the analyzed repositories. We assigned each repository to a FreeBSD ports-style category based on its primary purpose (as indicated by the repository description and documentation) and then mapped categories to a small set of descriptive domain tags (e.g., \texttt{www} $\rightarrow$ Web, \texttt{sysutils}/\texttt{net} $\rightarrow$ Systems/DevOps) to improve interpretability.

\subsection{File-Level Analysis}
An automated script processed each repository to identify files with high modification frequencies. First, files were categorized into programming files (source code) and administrative files or operational data (\eg documentation, configuration)~\cite[p. 241]{Hum89} based on their extensions and filenames.
Then, for each file, the script counted the number of commits in which the file was modified. Finally, statistical measures such as the mean and standard deviation of modification counts were calculated. 

We identified file-level hotspots through a dual-filter approach designed to minimize false positives while capturing genuine churn centers. Our method employed two complementary outlier detection criteria applied in conjunction. First, we flagged files whose modification counts exceeded three standard deviations above the mean ($\mu + 3\sigma$). Second, following the approach suggested by \citet{munaiah2017curating}, we required each file's total changes to exceed one change per month of project lifetime ($\texttt{lifetime\_months} \times 1$) to eliminate low-activity files from consideration. Only files satisfying both conditions were classified as potential hotspot hosts, as preliminary analysis revealed that either criterion applied independently produced excessive false positives. While tightening these thresholds---either by increasing the standard deviation cutoff or raising the monthly change rate---enhanced precision, it risked excluding legitimate centers of code churn.

\subsection{Line-Level Analysis}

To identify hotspots, we performed fine-grained line-level analysis on hotspot files. The task has two parts: (i) counting how often a concrete line changes and (ii) following that same line forward in time even as its position gets shifted by other changes/commits in the file.

We relied on Git to emit consecutive, patch-ordered
diffs:
\begin{verbatim}
$ git log -M -C --pretty=format:'commit %H %ct' \
         --reverse -p [file_path]
\end{verbatim}
The \verb|-M| flag enables move detection, allowing Git to track when code blocks are relocated.
The \verb|-C| flag enables copy detection, identifying when code is duplicated across files.
The \verb|--reverse| flag ensures commits are processed in chronological order from oldest to newest.
The \verb|-p| flag ensures that the diff is printed in patch format, which is useful for our line-level analysis.
\subsubsection{Line-Modification Counting Method}

Based on an existing peer-reviewed tool that determines the lifetime
of fine-grained elements by tracking specific lines across modifications~\cite{SLK21},
we developed a line change tracking algorithm
(\Cref{alg:line-modification}) that maintains, for every file, \texttt{file\_lines}, a vector of canonical line objects (including content, commit id, modification count, birth timestamp, \etc), and \texttt{line\_offsets}, a cumulative map from original to current indices.
When a modification appears in a line difference hunk reported
by Git, the algorithm:

\begin{enumerate}
\item adjusts the raw hunk coordinates by the running offset,
\item pairs deletions with additions when their positions align,
      copying the \textit{birth} timestamp and incrementing
      \textit{mod\_cnt}, and
\item records any unmatched deletion (resp.\ addition) as
      a \textit{death} (resp.\ new \textit{birth}).
\end{enumerate}
If a line is unchanged in successive commits, its lifespan continues
uninterrupted and its identifier (the slot in \texttt{file\_lines})
remains stable.

\paragraph{Line-identity model and offset adjustment.}
The \texttt{line\_offsets} map is what lets us ``follow'' the same logical line when earlier edits push it up or down.  After every processed hunk
we update the offset for all subsequent line indices.

\paragraph{Handling moved lines.}
We deliberately do not attempt to recognise intra-file moves. Once a block disappears from its original coordinates, the tracker marks those lines as dead; re-appearing identical text elsewhere is logged as newly added lines.  This positional-identity assumption keeps the algorithm lightweight and deterministic,
at the cost of missing some potential hotspots.

\begin{algorithm*}
      \caption{Git Diff Line Modification Tracker}
  \label{alg:line-modification}
  \begin{algorithmic}[1]
    \State \textbf{Initialize} data structures: file\_lines, line\_offsets, max\_processed\_index
    \State \textbf{Set} state $\gets$ ``commit''
    
    \While{not end-of-input}
      \If{state = ``commit''}
        \If{current\_commit exists}
          \State Finalize previous commit and reset tracking data
        \EndIf
        \If{line matches ``commit $<$hash$>$ $<$timestamp$>$``}
          \State Extract commit\_hash and timestamp
          \State Read next lines until encountering ``diff'', next ``commit'', or EOF
          \State Update state accordingly
        \EndIf
        
      \ElsIf{state = ``diff''}
        \If{line matches ``diff --git a/$<$old$>$ b/$<$new$>$''}
          \State Extract old\_file and new\_file paths
          \If{files are in target list}
            \State Initialize tracking for these files if needed
            \State Skip header lines until reaching ``@@ '' line
            \State state $\gets$ ``range''
          \Else
            \State Skip to next diff or commit
          \EndIf
        \EndIf
        
      \ElsIf{state = ``range''}
        \State Parse hunk header ``@@ -X,Y +A,B @@''
        \State Calculate adjusted file positions using line\_offsets
        \State Check for overlap with previously processed hunks
        \State Collect all lines in current hunk
        
        \State \textbf{Process hunk:}
        \Indent
          \State old\_idx $\gets$ old\_start
          \State updated\_lines $\gets$ []
          
          \For{each sequence of lines in hunk}
            \If{context line (starts with `` '')}
              \If{overlap $>$ 0}
                \State Skip and decrement overlap
              \Else
                \State Copy existing line to updated\_lines
              \EndIf
            \ElsIf{modification (sequence of ``-'' lines followed by ``+'' lines)}
              \State Group consecutive deletions and additions
              \For{each pair or unpaired line}
                \If{paired (modification)}
                  \State Create new line with updated content
                  \State Copy birth timestamp from old line
                  \State Increment mod\_count
                  \State Append to content, commit, and timestamp histories
                \ElsIf{deletion only}
                  \State Mark old line with death\_ts = current timestamp
                \ElsIf{addition only}
                  \State Create new line with birth\_ts = current timestamp
                \EndIf
              \EndFor
            \EndIf
          \EndFor
        \EndIndent
        
        \State Replace affected lines in file\_lines with updated\_lines
        \State Update max\_processed\_index to track processed areas
        \State Determine next state based on next line
        
      \ElsIf{state = ``EOF''}
        \State Break main loop
      \EndIf
      
      \State Read next line if needed
    \EndWhile
    
    \State \textbf{Output results:}
        \For{each line info in file}
          \If{line has no death timestamp (still alive)}
            \State Write to CSV: line number, content, mod\_count, and history details
          \EndIf
        \EndFor
  \end{algorithmic}
\end{algorithm*}

\subsection{Taxonomy Development}
To develop a taxonomy of hotspots, we employed a systematic manual labeling process complemented by similarity detection techniques. This approach ensured that the taxonomy accurately reflects the diverse types of line modifications observed in the data.

\subsubsection{Manual Labeling Process}
Two of this paper's authors independently reviewed the files generated from line change data.
For each hotspot, both researchers first documented their individual interpretations of what was causing the changes, including their reasoning and observations. After this independent documentation phase, the researchers convened to compare their findings, discuss their reasoning, and assign an appropriate tag that best described the nature of the modification. This two-step process---individual documentation followed by collaborative convergence---ensured both independent analysis and consistency in the final labeling, following established systematic review practices where reviewers work independently before resolving disagreements through discussion~\cite{brereton2007lessons}.

To determine labeling saturation, we applied the Chao1 estimator~\cite{smith1984nonparametric},
which has been adopted in several prior software engineering studies~\cite{miller1999estimating, bohme2018stads, liyanage2023reachable}.
It is a non-parametric lower-bound estimator of species (here hotspot-types)
richness and is defined as:

\begin{equation}
S_{\text{est}} = S_{\text{obs}} + \frac{f_1^2}{2f_2}
\end{equation}

In this context, \(S_{\mathrm{obs}}\) is the number of distinct hotspot types actually observed in our sample, \(f_{1}\) is the number of hotspot types seen exactly once (singletons), and \(f_{2}\) is the number of hotspot types seen exactly twice (doubletons).

\begin{figure}[t]
  \centering
  \setlength{\fboxsep}{0pt}
    \includegraphics[width=\linewidth, trim=0 0 0 0, clip]{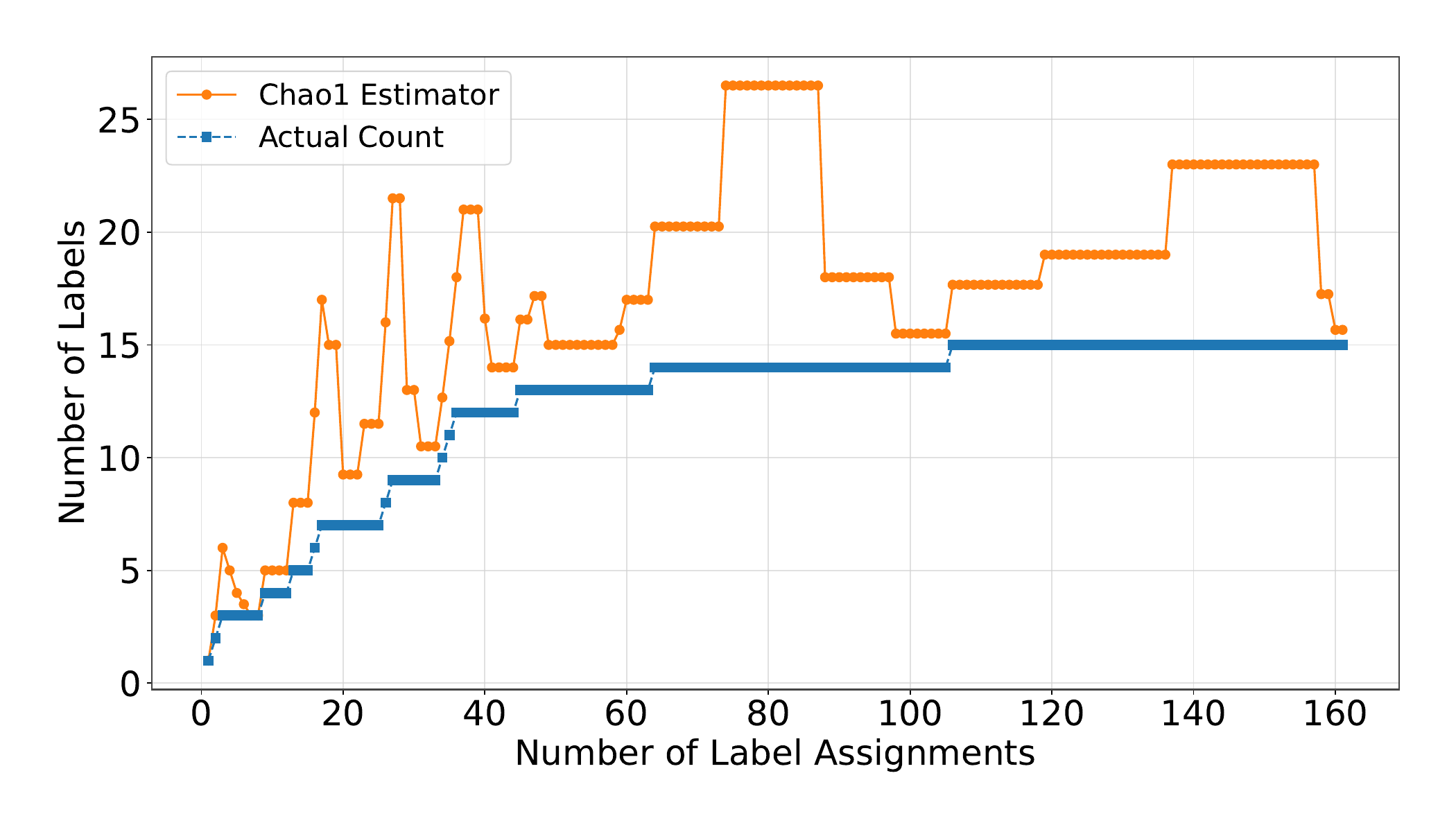}
  \caption{Chao1 Estimator vs. Number of Label Assignments}
  \label{fig:chao1}
\end{figure}

In Figure~\ref{fig:chao1}, the Chao1 estimate rises quickly as new hotspot types appear, then levels off, showing diminishing returns from further labeling. After about 100 assignments the observed count stabilizes
at 15 unique hotspot types and stays flat through 160.
The Chao1 lower bound remains slightly above this and ends at 15.67 at 160 assignments, a difference of about 0.67 (approximately 4.4 percent). This plateau indicates that the sample is saturated, so additional labeling would likely uncover at most one more label at disproportionate cost.
This statistical approach provided a systematic stopping criterion for our manual labeling process, allowing us to make a reasoned judgment about when we had identified a representative sample of the hotspot types present in our dataset.

We manually labeled and investigated the mined hotspots, allowing us to label only a fraction of our mined ones until the number of defined labels reached
the goal established by the Chao1 estimator.
Our experiment did not aim to label all hotspots;
instead, our goal was to describe all major hotspot types.

\begin{table}[t]
  \centering
  \caption{Repository Domain Distribution}
  \label{tab:repo-categories}
  \begin{tabular}{l l r}
  \toprule
  Domain tag & FreeBSD category & Count \\
  \midrule
  Web & www & 19 \\
  Systems/DevOps & sysutils & 14 \\
  Dev tools \& SDKs & devel & 10 \\
  Miscellaneous & misc & 7 \\
  Games & games & 6 \\
  Science/Engineering & science & 5 \\
  Python Packages & python & 4 \\
  Languages & lang & 4 \\
  Mathematics & math & 3 \\
  Systems/DevOps & net & 3 \\
  Docs/Content & textproc & 3 \\
  Dev tools & editors & 3 \\
  Java Packages & java & 2 \\
  Graphics & graphics & 2 \\
  Databases & databases & 2 \\
  Desktop/UI & x11 & 2 \\
  Finance/Crypto & finance & 1 \\
  Security & security & 1 \\
  \bottomrule
  \end{tabular}
  \end{table}
  
After labeling, the results were consolidated. If both researchers assigned the same label, that label was considered final. If the labels differed, the researchers discussed their reasoning and decided on a label accordingly. If they were unable to reach a consensus,
the paper's last author --- a leading (R4) researcher~\cite{FGK15}
with considerable software engineering expertise --- was consulted,
and a tag was mutually decided upon.
The inter-rater agreement of the two initial software engineering researchers using Cohen's Kappa was found to be 0.61, which denotes a substantial level of agreement between the two raters~\cite{landis1977measurement}.

\subsection{Bot Detection}
To identify bots that contributed to hotspots, we analyzed commit data. For each hotspot line, we aggregated the commit count made by each contributor. This was achieved by grouping commits based on the committer's e-mail and name, which we extracted using the \texttt{git show} command on the relevant commit IDs.

We consolidated all committer information into a single file. To do this, we grouped contributors with the same name and e-mail, then applied case-insensitive regular expressions to flag usernames or display names containing keywords like ``bot'' or ``auto.'' To ensure high accuracy, we manually reviewed every flagged entry and identified just one false positive---``Drobotov,'' whose name inadvertently matched our filter. This process allowed us to pinpoint genuine bots and assess their significant impact on code-churn.

By aggregating contributor names or e-mails associated with hotspots across all projects and analyzing the most frequent ones, we could effectively identify and isolate bot activity.

\section{Results}
\label{sec:results}
This section presents our identified hotspot types, their characteristics, and the role of automated contributions in generating these patterns.

\subsection{Identified Hotspot Types (RQ1)}
Through our analysis, we discovered and characterized a total of 15 unique hotspot types, which have been divided into four umbrella categories. Each hotspot type is described below, along with a representative example demonstrating the changes' nature.
All colored code lines are hyperlinked
to permanent Software Heritage archive~\cite{DZ17} locations
using the ISO/IEC 18670:2025 standard for SoftWare Hash IDentifiers
(SWHIDs)~\cite{ISO18670:2025}.
We also identified and classified instances of normal \emph{software evolution}
which are legitimate and can appear as frequent edits but do not represent anti-patterns.

\subsubsection{Configuration Management Hotspots}
These include changes to version specifications, dependencies, and resource configurations that govern software component interactions and deployment.

\PatternItem{Pinned Version Bump} {involves incrementing a fixed (pinned) version in the absence of flexible or semantic versioning strategies. }

\beforeafter{%
\href{https://archive.softwareheritage.org/swh:1:cnt:7a88a47aedc7ff48f9d6e10efe0042b9204f2187;anchor=swh:1:rev:cebbe734822da36aee2ecc9c640b6e792af65a80/\#swh-revision-changes}{release = "3.10.181"}%
}{%
\href{https://archive.softwareheritage.org/swh:1:cnt:7a88a47aedc7ff48f9d6e10efe0042b9204f2187;anchor=swh:1:rev:cebbe734822da36aee2ecc9c640b6e792af65a80/\#swh-revision-changes}{release = "3.10.182"}%
}

Frequent changes in pinned versions can be avoided by using version ranges instead of fixed version pins. We also suggest avoiding committing files which contain locked versions (\eg \texttt{pipfile.lock}, \texttt{package-lock.json}). Semantic versioning can also help us here by updating versions selectively when there's a meaningful change (bug fix, feature, or breaking change). This contrasts with arbitrary version bumping that might happen on every release cycle regardless of significance.

\PatternItem{Conditional Version Bump}{involves adjusting conditions related to version checks. }

\beforeafter{%
\href{https://archive.softwareheritage.org/swh:1:cnt:3e2e6379273893ce408c6b3dabdfd272a96ee348;anchor=swh:1:rev:83807088329b2a7e6422e0d0ba460870a265d3d2/\#diff_0aa38970b36a524f9ee36f29088ecfb1a58b74f0+F4T0-F4T0-unified}{numpy>=1.15,<1.19.0}%
}{%
\href{https://archive.softwareheritage.org/swh:1:cnt:3e2e6379273893ce408c6b3dabdfd272a96ee348;anchor=swh:1:rev:83807088329b2a7e6422e0d0ba460870a265d3d2/\#diff_0aa38970b36a524f9ee36f29088ecfb1a58b74f0+F0T4-F0T4-unified}{numpy>=1.16.5,<1.19.0}%
}

The same mitigation strategies as for Pinned Version Bump can be followed to avoid frequent conditional version bumps.

\PatternItem{Resource ID Modification}
{involves changes to resource identifiers such as Docker images or Amazon Machine Image (AMI) IDs. Small oversights in these references can lead to deploying incorrect or deprecated resources.}

\beforeafter{%
\href{https://archive.softwareheritage.org/swh:1:cnt:ab7804bcfae1a569e5e9a682bf360cbc380f8a66;anchor=swh:1:rev:ba8b4518c16fdf4c0aad2cd956f2cbb0be68fc88/\#diff_2e24ea334178e6bd9c56d8c1283ea4cbe413a0cf+F29T0-F29T0-unified}{IMAGE=container-vm-v20141208}%
}{%
\href{https://archive.softwareheritage.org/swh:1:cnt:ab7804bcfae1a569e5e9a682bf360cbc380f8a66;anchor=swh:1:rev:ba8b4518c16fdf4c0aad2cd956f2cbb0be68fc88/\#diff_2e24ea334178e6bd9c56d8c1283ea4cbe413a0cf+F0T29-F0T29-unified}{IMAGE=container-vm-v20150112}%
}

To minimize these changes, provide the resource IDs as command-line parameters, or retrieve them through APIs using stable tags, or store them in environment variables.

\PatternItem{Service Configuration}{involves modifications to service configuration, including SSH tokens, IP addresses, and port numbers.}

\beforeaftermultiline{%
\href{https://archive.softwareheritage.org/swh:1:cnt:b0b23d8ecf0ff93c7226df2c0c47efbfb39bff81;anchor=swh:1:rev:ec0b1162504eae24edf788d0301fabae58dde8d9;path=ansible/inventory/env/group_vars/all.yml;lines=201}{sunbird\_user\_service\_api\_base\_url: ``http://\{\{sunbird\_swarm\_manager\_lb\_ip\}\}:9000''}%
}{%
\href{https://archive.softwareheritage.org/swh:1:cnt:b0b23d8ecf0ff93c7226df2c0c47efbfb39bff81;anchor=swh:1:rev:2c0b6b319fa5d160e2cd656c81894313238b36bf;path=ansible/inventory/env/group_vars/all.yml;lines=201}{sunbird\_user\_service\_api\_base\_url: ``http://\{\{private\_ingressgateway\_ip\}\}/learner''}%
}

To avoid these changes, move external service configurations to environment variables or command-line arguments.
This also prevents sensitive information from appearing in hotspots. It is also suggested to use tools like  Terraform, Ansible, or Kubernetes with secret management solutions (HashiCorp Vault, AWS Secrets Manager, \etc) and define configurations declaratively.

\PatternItem{Dependency Specification}
 {involves adjustments in dependency declarations, such as adding, removing, or modifying modules in import statements or dependency arrays, or updating entries in a \texttt{package.json}, \texttt{pom.xml}, or similar file.}

 \beforeaftermultiline{%
 \href{https://archive.softwareheritage.org/swh:1:cnt:399ff510d79c25caaa22dc2f51b64487cf3ade5e;anchor=swh:1:rev:d1f81da2361804d62f7ade2703eda68dda6cdb48/\#diff_5c1d3ed13af75459f297ec1dbb964a7d309d4e4d+F13T0-F13T0-unified}{obj-\$(CONFIG\_VIDEO\_DEV) += videodev.o compat\_ioctl32.o v4l2-int-device.o}%
 }{%
 \href{https://archive.softwareheritage.org/swh:1:cnt:399ff510d79c25caaa22dc2f51b64487cf3ade5e;anchor=swh:1:rev:d1f81da2361804d62f7ade2703eda68dda6cdb48/\#diff_5c1d3ed13af75459f297ec1dbb964a7d309d4e4d+F0T13-F0T13-unified}{obj-\$(CONFIG\_VIDEO\_DEV) += videodev.o v4l2-compat-ioctl32.o v4l2-int-device.o}%
 }
 
 To avoid frequent dependency specification changes, implement a robust dependency management strategy. 
 Automated dependency update tools (e.g., Dependabot, Renovate) can reduce manual version edits by systematically batching and standardizing dependency updates. However, without appropriate configuration, such tools may generate excessive update churn through overly frequent or fragmented updates. Controlled update policies (e.g., scheduled batched updates, grouped dependency bumps) can mitigate this risk while maintaining security currency.

 \PatternItem{External Data Fluctuations}
{
are driven by external data sources, such as including updated currency exchange rates in a database or regenerating code from an external API specification. 
}

\beforeafter{%
\href{https://archive.softwareheritage.org/swh:1:cnt:04a3ca745b1ba80280375267b4c045293efb1015;anchor=swh:1:rev:48eeb35e4b253fcd136952b04d9744b272a8fd7e;path=data/formats-data.ts;lines=479}{tier: ``NU'',}%
}{%
\href{https://archive.softwareheritage.org/swh:1:cnt:04a3ca745b1ba80280375267b4c045293efb1015;anchor=swh:1:rev:b9adeafd1ed327b4565cbe71ef7d582ec69bfedd;path=data/formats-data.ts;lines=466}{tier: ``PU'',}%
}

To remove external data fluctuations,
obtain the corresponding data automatically from stable interfaces
for external data sources.
The performance impact cab be minimized by employing caching mechanisms that
reduce the update frequency and impact.

\subsubsection{Development Environment Hotspots}
These include modifications to development infrastructure including file paths, base distributions, and debugging settings that affect the build environment.

\PatternItem{Path Update}{occurs when file paths or imports are altered, often due to project restructuring or changes in directory organization. This also includes modifications made to URL links.}

\beforeafter{%
\href{https://archive.softwareheritage.org/swh:1:cnt:542af85ba2bb85fe5f8a9df0dd8fac6b01abd49d;anchor=swh:1:rev:897e97330a3fc5e25f700fa2aa86edb9ee5cbc21;path=src/lib/db/project-stats-store.ts;lines=7}{import \{ IProjectStats \} from `lib/services/project-service';}%
}{%
\href{https://archive.softwareheritage.org/swh:1:cnt:542af85ba2bb85fe5f8a9df0dd8fac6b01abd49d;anchor=swh:1:rev:1392b10727967b14cbb9578e5f5c9900cdab4310;path=src/lib/db/project-stats-store.ts;lines=6}{import \{ IProjectStats \} from `../services/project-service';}%
}

To avoid frequent path updates, it is suggested to structure the project according to established guidelines such as those by Maven. In addition, projects should employ \emph{stable import aliases} that decouple logical module names from their physical directory layout. Mechanisms such as TypeScript path mappings, bundler aliases (e.g., Webpack or Vite), or equivalent language-specific solutions allow directory reorganization without requiring widespread import rewrites. Abstracting imported resources and URL specifications into configuration files further minimizes churn and reduces the risk of broken imports during structural changes.

\PatternItem{Distro Bump}{refers to upgrading or changing the base distribution (\eg Linux distribution, container base image) to a newer release. }

\beforeafter{%
\href{https://archive.softwareheritage.org/swh:1:cnt:2dabf86f9a0e192d0df0e842429f4b568a0897f1;anchor=swh:1:rev:43f68d948ef98917972419b7ed00b811214dbc0f;path=zuul.d/jobs.yaml;lines=40}{name: kolla-ansible-centos8s-source-kvm}%
}{%
\href{https://archive.softwareheritage.org/swh:1:cnt:2dabf86f9a0e192d0df0e842429f4b568a0897f1;anchor=swh:1:rev:67607c679e20adaa42bc5271110a6b0aa992ffc8;path=zuul.d/jobs.yaml;lines=54}{name: kolla-ansible-rocky9-source-kvm }%
}

To mitigate the need for frequent distro bumps, choose a stable and long-term supported base distribution.

\PatternItem{Debug Configuration}{
refers to changes in debugging or logging settings (\eg adding \texttt{-v} or \texttt{-r} flags to commands). While often required for troubleshooting, leaving excessive debugging configurations in production can introduce overhead or leak sensitive data. }

\beforeaftermultiline{%
\href{https://archive.softwareheritage.org/swh:1:cnt:c2a7857742911e55b3fe3e222db6c4b108f9b55e;anchor=swh:1:rev:685d5f272284ae7a82396837ee42f98bd99d7211/\#diff_6108307980ae51a2927f69c5ae89c1d87450ed12+F325T0-F325T0-unified}{extra\_var\_arg+=` -e instance\_userdata=``'' -e launch\_wait\_time=0'}%
}{%
\href{https://archive.softwareheritage.org/swh:1:cnt:c2a7857742911e55b3fe3e222db6c4b108f9b55e;anchor=swh:1:rev:685d5f272284ae7a82396837ee42f98bd99d7211/\#diff_6108307980ae51a2927f69c5ae89c1d87450ed12+F0T325-F0T325-unified}{extra\_var\_arg+=` -e instance\_userdata=``'' -e launch\_wait\_time=0 -e elb\_pre\_post=false'}%
}

To avoid frequent debug configuration changes, debugging and logging parameters should be externalized into environment variables or feature-flag frameworks rather than committed directly to source files. Tools such as environment-variable loaders can support this separation, while continuous-integration checks can detect and flag accidental commits of debug-specific settings, preventing unnecessary churn.

\subsubsection{Code Quality and Style Hotspots}
These include formatting adjustments, function signature changes, and structural improvements that enhance code readability without altering functionality. \\

\PatternItem{Function Call Change}{encompasses alterations in function signatures (parameters, return types) or renaming of method calls.}

\beforeaftermultiline{%
\href{https://archive.softwareheritage.org/swh:1:cnt:1d05dc1c7010d58039893dea81ffc0346d66968e;anchor=swh:1:rev:ed1c9977cb1b63e4270ad8bdf967a2d02580aa08/\#diff_334358e4a72ed8fa9674d6d0bdc708cf2cf25849+F168T0-F168T0-unified}{printf(``\%s'', find\_unique\_abbrev(get\_object\_hash(parent->object), abbrev));}%
}{%
\href{https://archive.softwareheritage.org/swh:1:cnt:1d05dc1c7010d58039893dea81ffc0346d66968e;anchor=swh:1:rev:ed1c9977cb1b63e4270ad8bdf967a2d02580aa08/\#diff_334358e4a72ed8fa9674d6d0bdc708cf2cf25849+F0T168-F0T168-unified}{printf(``\%s'', find\_unique\_abbrev(parent->object.oid.hash, abbrev));}%
}

To minimize the need for frequent function call changes, define stable interfaces for core functions and create stable wrapper functions or adapters.
In languages that support abstract classes and interfaces,
(such as Java, C\#, TypeScript, and Kotlin),
follow the robustness principle~\cite{All11} by
passing arguments as abstract class types
and returning values as interface types.
In languages that support keyword argument passing,
(such as Python, Ruby, Kotlin, Swift, Scala, Julia, and Elixir),
employ named rather than positional arguments.
In most object-oriented languages pass arguments as object,
use the builder pattern~\cite[p. 97]{GHJV95}, or employ polymorphism.

\PatternItem{Formatting Ping-Pong}{represents purely stylistic or formatting-related changes, such as adjusting whitespace, using tabs, or toggling between uppercase and lowercase. These changes can clutter commit histories and obscure more critical modifications.}

\beforeafter{%
\href{https://archive.softwareheritage.org/swh:1:cnt:a7a3b8e8a03e71db766e5454b981d22a1299711b;anchor=swh:1:rev:c37742de0eb972bdeced44b4a37502a4fae4101e/\#diff_85427d60e76377714ffe938ebfceba3995161c54+F44T0-F44T0-unified}{\#include <assert.h>}%
}{%
\href{https://archive.softwareheritage.org/swh:1:cnt:a7a3b8e8a03e71db766e5454b981d22a1299711b;anchor=swh:1:rev:c37742de0eb972bdeced44b4a37502a4fae4101e/\#diff_85427d60e76377714ffe938ebfceba3995161c54+F0T44-F0T44-unified}{\#include <assert.h>}%
}

To minimize Formatting Ping-Pong, automated linters and code formatters (\eg Prettier, Black, gofmt, clang-format, \etc) should be used. CI/CD pipelines can also be configured to validate formatting compliance with an agreed upon standard.

\PatternItem{Long Line Change}{captures edits to long lines of text or code, such as extensive comments, documentation paragraphs, long import statements, or expressions.
The length of modified lines often obscures the actual changes in the output of common line-oriented diff algorithms.}

\beforeaftermultiline{%
\href{https://archive.softwareheritage.org/swh:1:cnt:0b1c8084ec4c402dc11dc5e26f6345b30462a4bf;anchor=swh:1:rev:c0f5fcb48437e4d3c148bd2742a9fc92eb49630a/\#diff_0a2a121401098d9ad1806626210e48f049841607+F164T0-F164T0-unified}{On the right side of the canvas is Search, and the Global Menu. You can use Search to easily find components on the}%
}{%
\href{https://archive.softwareheritage.org/swh:1:cnt:0b1c8084ec4c402dc11dc5e26f6345b30462a4bf;anchor=swh:1:rev:c0f5fcb48437e4d3c148bd2742a9fc92eb49630a/\#diff_0a2a121401098d9ad1806626210e48f049841607+F0T164-F0T164-unified}{On the right side of the canvas  is Search, and the Global Menu. For  more information on search refer to <<search>>. The Global Menu}%
}

To minimize long line changes, break down long lines into smaller ones. This can include refactoring long lines of code into smaller functions or methods.
In the documentation, employ \emph{semantic line breaks}~\cite[p. 11]{Ker79},
starting each phrase in a new line and breaking long lines near 60--70 columns.
Linting rules should be implemented that flag when strings exceed a certain length. We also suggest creating pre-commit hooks to enforce these practices.

\subsubsection{Administrative Hotspots}
These include updates to non-functional metadata and documentation for legal compliance and project management.

\PatternItem{License Modification}{refers to updates in copyright headers, entity names, years, or licensing text. While these changes are often boilerplate, they need to be minimized or performed accurately to maintain legal clarity.}
 
\beforeafter{%
\href{https://archive.softwareheritage.org/swh:1:cnt:cc306ab0ce120d3d5e92d09b33b5c5442a303b6c;anchor=swh:1:rev:20004050bcd9396f504e3e33138b734d96de5238/\#diff_dc66a87ddedc29515d2c45e62530c9185b5cd7bb+F1T0-F1T0-unified}{\# Copyright (C) 2008-2013 TrinityCore <http://www.trinitycore.org/>}%
}{%
\href{https://archive.softwareheritage.org/swh:1:cnt:cc306ab0ce120d3d5e92d09b33b5c5442a303b6c;anchor=swh:1:rev:20004050bcd9396f504e3e33138b734d96de5238/\#diff_dc66a87ddedc29515d2c45e62530c9185b5cd7bb+F0T1-F0T1-unified}{\# Copyright (C) 2008-2014 TrinityCore <http://www.trinitycore.org/>}%
}
 
To avoid frequent license modifications, simplify license maintaince processes, \eg by not updating the copyright year. Rather than having the whole license in each file, only provide a reference to a single shared license file.

\PatternItem{Metadata Change}
{involves updating automatically generated information such as timestamps, signatures, checksums, or version identifiers. These changes can be misleading if included in commits without actual content changes, as they inflate the perceived scale of a revision. }

\beforeafter{%
\href{https://archive.softwareheritage.org/swh:1:cnt:f0168c17151b74c23acb57181c42f09b58da1f14;anchor=swh:1:rev:c850c6f2e97adf2fc8f46844a7f8aa70d2d4ba5e/\#diff_25cc56e78417d300b8008dbae6fc7250381e6727+F4T0-F4T0-unified}{``timestamp'': ``2020-10-03T12:37:57.000+00:00'',}%
}{%
\href{https://archive.softwareheritage.org/swh:1:cnt:f0168c17151b74c23acb57181c42f09b58da1f14;anchor=swh:1:rev:c850c6f2e97adf2fc8f46844a7f8aa70d2d4ba5e/\#diff_25cc56e78417d300b8008dbae6fc7250381e6727+F0T4-F0T4-unified}{``timestamp'': ``2021-01-25T14:24:58.697Z'',}%
}

To avoid frequent metadata changes, reduce the amount of metadata stored
under revision control by automating their generation during the build process.
Use build tools like Maven, Gradle, or setuptools-scm to derive version information from Git tags rather
than hard coding metadata in source code files.
Adopting the goal of reproducible software builds~\cite{LZ22},
will reduce needless metadata and thus corresponding changes,
while also increasing the software's trustworthiness.

\PatternItem{Stepwise Refactoring}
{reflects code modifications intended to restructure or improve readability without affecting the external behavior. The modifications made in this hotspot type are closer in time. While code refactoring is essential, frequent small-scale changes may clutter the commit history, making it harder to track significant modifications. }

\beforeaftermultiline{%
\href{https://archive.softwareheritage.org/swh:1:cnt:02aed119c1152d218619daf149152d6fe00b66fe;anchor=swh:1:rev:5f1e47254edc1d69a0a928e2bd7d9a6f64c21f9e/\#diff_5c32b7c67795c67d4dace52ef8c1296f11f9b11a+F1302T0-F1302T0-unified}{menu\_opts = model\_title.starts\_with?(``Parent'') ? \{\} : \{:menu => chart[:menu], :zoom\_url => zoom\_url\}}%
}{%
\href{https://archive.softwareheritage.org/swh:1:cnt:02aed119c1152d218619daf149152d6fe00b66fe;anchor=swh:1:rev:5f1e47254edc1d69a0a928e2bd7d9a6f64c21f9e/\#diff_5c32b7c67795c67d4dace52ef8c1296f11f9b11a+F0T1302-F0T1302-unified}{menu\_opts = parent ? \{\} : \{:menu => chart[:menu], :zoom\_url => zoom\_url\}}%
}

Stepwise refactoring is often necessary to safely address architectural and design problems. Hotspots arise when refactoring proceeds through many small, mechanically repetitive edits to the same lines, inflating churn without providing globally useful intermediate states. To reduce version-history noise while preserving safe refactoring practices, developers can (i) maintain explicit refactoring plans with defined invariants, (ii) use short-lived branches and squash intermediate commits before merging, and (iii) employ automated refactoring tools where available. This mitigation targets version-history noise rather than discouraging incremental refactoring itself.
\subsection{Hotspot Characteristics (RQ2)}
\begin{figure}[b]
  \centering
  \includegraphics[width=\linewidth]{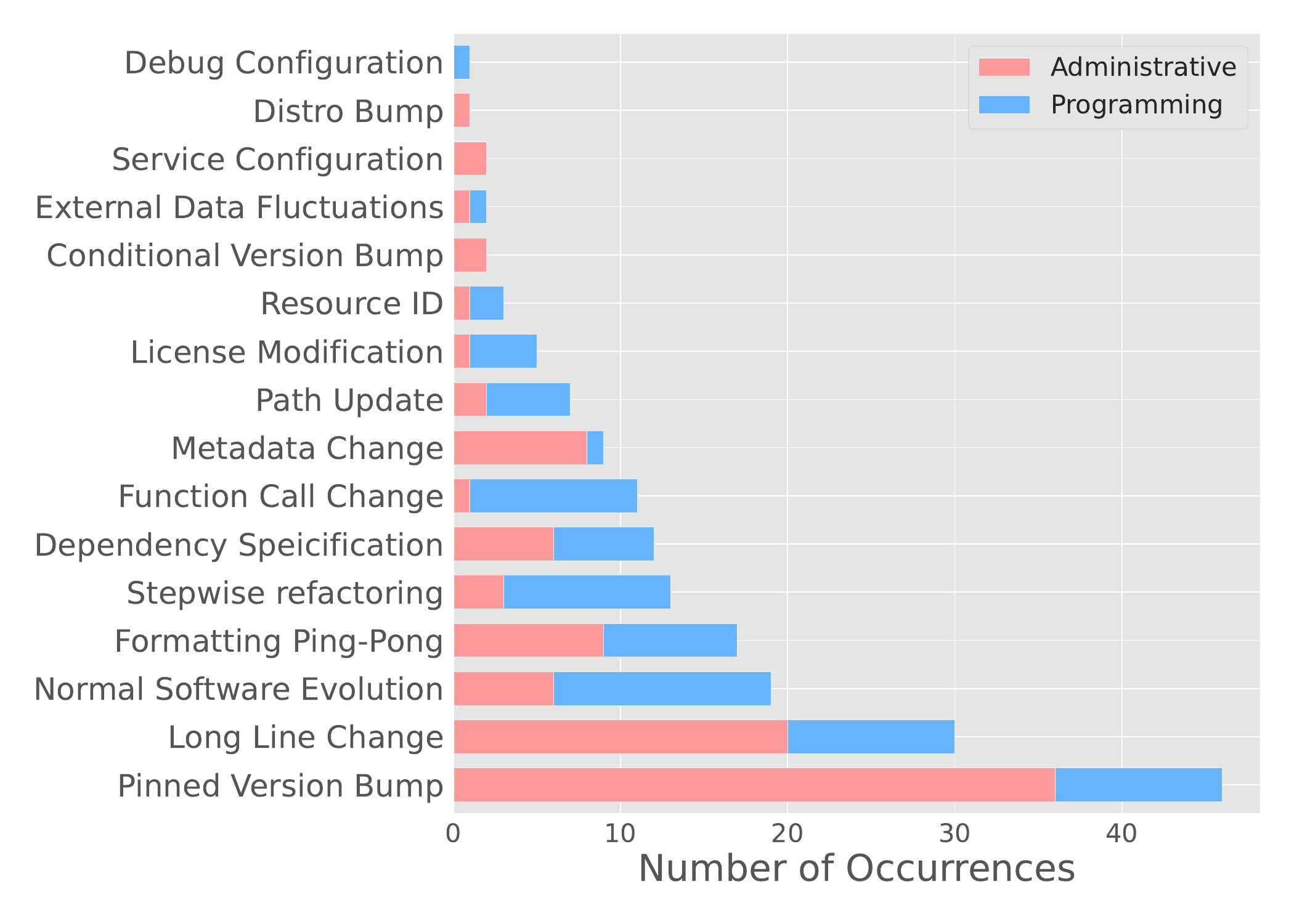}
  \caption{Distribution of Occurrences per Hotspot Type}
  \label{fig:distribution}
\end{figure}

Our analysis indicates a significant distribution between changes in administrative and programming files across the identified hotspots
(see \Cref{fig:distribution}). Some repositories only had hotspots in administrative files and not in the source-code files. Therefore, administrative changes account for 53.6\% of all the identified hotspots.

Many hotspot types stem from administrative files: \emph{Pinned Version Bump}, \emph{Conditional Version Bump}, \emph{Long Line Change}, and \emph{Metadata Change}. In contrast, \emph{Function Call Change}, \emph{Debug Configuration}, \emph{Resource ID Modification}, and \emph{Path Update} show a stronger programming orientation.  This split highlights how understanding the nature of changes can guide different intervention strategies. For example, addressing administrative hotspots may require improved configuration management systems, while programming-dominated hotspots might benefit from design and architectural refactoring (e.g., interface stabilization and modularization) to localize change.

~\Cref{tab:hotspot-baseline} shows that hotspot activity exhibits extreme skewness across our dataset. Among 91 repositories, the total file count was 3\,277\,296, with 28\,470 outlier files (0.87\% of all files) containing at least one hotspot line. This indicates that hotspots are relatively rare, yet they can have a significant impact on the overall codebase. Among 174 randomly selected hotspot-bearing files, we identified 4\,052 outlier lines with a highly skewed distribution: while the median file contains only seven outlier lines, the mean reaches 23.3, and one extreme file harbors 491 outliers. Analysis of outlier line lifespans (measured as the temporal difference between first and last modifications) demonstrates remarkable longevity: half persist for $\geq$1\,198 days ($\approx$3.3 years), the upper quartile survives beyond 2\,064 days (5.6 years), and the most persistent hotspot endured 18.1 years. Regarding editing intensity, the median hotspot undergoes eleven modifications, yet exhibits heavy-tailed behavior with 25\% of hotspots exceeding 37 edits and the most pathological line experiencing 1\,864 changes. 
These empirical findings substantiate our hypothesis that a small subset of lines concentrates a disproportionate share of long-term maintenance effort, thereby warranting specialized treatment in our analytical framework.

\begin{table}[t]
  \centering
  \caption{Descriptive statistics for hotspots
           (4\,052 lines drawn from 174 files).}
  \label{tab:hotspot-baseline}
  \begin{tabular}{@{}lrrrrr@{}}
    \toprule
    & Min & Median & Mean & Max & IQR \\
    \midrule
    Hotspot lines/file                & 1.00   & 7.00     & 23.29 & 491.0  & 21.75 \\
    Lifespan (years)                 & 0.00 & 3.28 & 4.29 & 18.13 & 3.83 \\
    Modification Count                & 3.00   & 11.0    & 87.40 & 1\,864 & 32.0 \\
    \bottomrule
  \end{tabular}
\end{table}

\subsection{Bot Activity (RQ3)}

Our analysis identified automated bots as significant contributors to hotspots across the studied projects. In total, 23 unique bots were identified (representing 2.8\% of the 825 total hotspot committers across all studied projects).
Table~\ref{tab:bots} lists the ten most active bots, with \textit{skia-flutter-autoroll} being the most prolific, responsible for 10\,541 commits. Collectively, these bots accounted for 15\,283 out of 20\,690 total analyzed commits (approximately 73.9\%).

\begin{figure}[b]
  \centering
  \includegraphics[width=\linewidth]{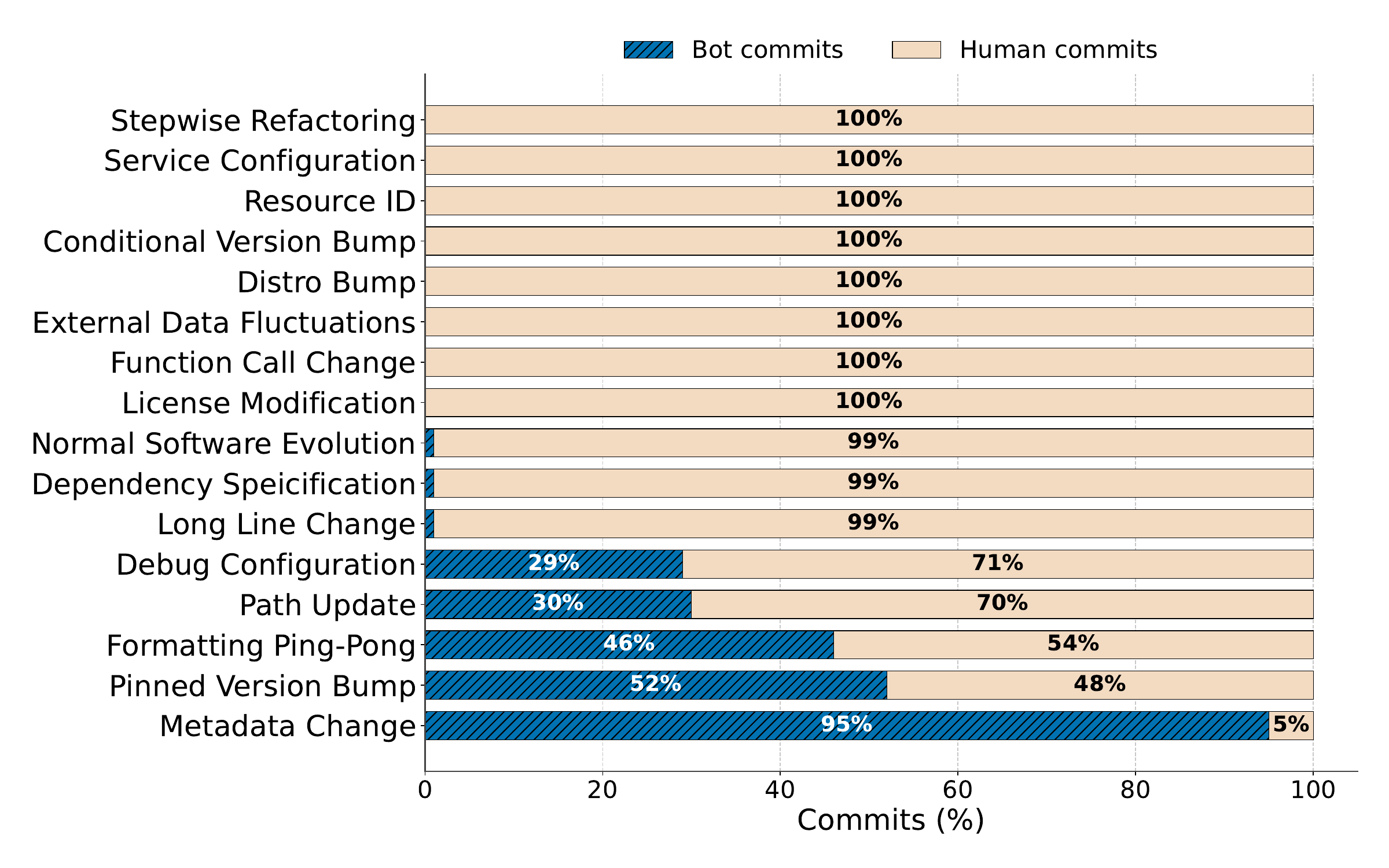}
  \caption{Bot vs. Human Commit Ratio}
  \label{fig:bot_vs_human}
\end{figure}

~\Cref{fig:bot_vs_human} depicts a distribution of automation across hotspot categories.  
Five patterns (\emph{Stepwise Refactoring}, \emph{Service Configuration}, \emph{Resource ID}, \emph{Conditional Version Bump}, and \emph{Distro Bump}) are entirely human-driven (100\% human commits).  
Three more (\emph{Normal Software Evolution}, \emph{Dependency Specification}, and \emph{Long Line Change}) show only small traces bot activity (\(<1\%\)).  
Conversely, bots generate nearly a third of all \emph{Debug Configuration} (29\%) and \emph{Path Update} edits (30\%). \emph{Formatting Ping-Pong} and \emph{Pinned Version Bump} are almost an even split (46\% and 52\% bot commits, respectively).  
The most extreme case is the \emph{Metadata Change}, where bots account for 95\% of the edits.  

This pattern confirms that currently bot automation concentrates on mechanically repetitive tasks (semantic-version bumps, formatting tweaks, and metadata refreshes), while changes that require domain knowledge (\eg function-call refactoring or external-resource updates) remain overwhelmingly human.  
From a maintainability standpoint, these findings suggest that maintenance
efforts should focus on
(i) suppressing bot-generated noise in code-review workflows and
(ii) highlighting human-authored hotspots
(which still dominate many error-prone areas)
as the primary targets for reviewing and refactoring.

The substantial proportion of bot-driven commits underscores their pivotal role in generating frequent, often repetitive changes. This finding aligns with prior research emphasizing the influence of bots on code churn \cite{dey2020exploratory}. For academia, these results highlight the necessity of differentiating automated contributions from human-driven development efforts when conducting repository analyses. In industry, recognizing the impact of bots can guide more effective metric interpretations, resource allocation, and optimization of automated workflows. Open-source software communities benefit by understanding these patterns, enabling maintainers and developers to fine-tune bot behaviors, mitigate unnecessary churn, and focus human attention on meaningful software development activities.

\begin{table}[t!]
\caption{Top Ten Identified Bots}
\centering
\begin{tabular}{l r}
\toprule
\textbf{Bot Name} & \textbf{Commit Count} \\
\midrule
skia-flutter-autoroll & 10541 \\
vercel-release-bot & 1044 \\
Electron Bot & 795 \\
dependabot[bot] & 530 \\
jenkins-x-bot & 483 \\
GitHub Actions Bot / github-actions[bot] & 441 \\
Sudowoodo Release Bot & 438 \\
Confluent Jenkins Bot & 130 \\
Protobuf Team Bot & 128 \\
Netty Project Bot & 124 \\
\bottomrule
\end{tabular}
\label{tab:bots}
\end{table}

\section{Discussion}
\label{sec:discussion}
This section interprets our results through the lens of the three research questions (RQs) and the maintainability dimensions of \emph{configurability}, \emph{stability}, and \emph{changeability}.
We also discuss implications for practitioners and future research.

\subsection{What factors give rise to hotspots and how can they be curbed? (RQ1)}

Our taxonomy uncovered fifteen recurring patterns.  Three of them
explain almost half of all hotspot edits:
\emph{Pinned Version Bump},
\emph{Long Line Change}, and
\emph{Formatting Ping-Pong}.
All three stem from tooling or process deficiencies rather than inherent domain volatility, confirming that a significant portion of hotspots is preventable. These patterns suggest three key principles for reducing hotspots:

\begin{description}
  \item[Configurability]  Pinned or conditional version bumps and resource-ID tweaks surface because values are committed rather than externalised. Moving these to declarative manifests or environment variables would eliminate the need for code edits.
  \item[Stability]  Frequent path rewrites and distro bumps indicate fragile directory structures and infrastructure baselines. Adopting stable import aliases and long-term-support base images would dampen ripple effects.
  \item[Changeability]  Formatting oscillations and long-line rewrites clutter histories and reviews. Enforcing automated formatters in CI avoids human ``ping-pong''.
\end{description}

These observations translate directly into actionable guidelines as provided in our paper.

\subsection{How do hotspots vary in scope, persistence, and location? (RQ2)}

Hotspot activity is \emph{extremely} skewed: a median of seven outlier lines per file contrasts with a mean of 23.3 and a maximum of 491 (\Cref{tab:hotspot-baseline}).  Lifespans are long (median 3.3 years; max 18.1 years) and editing intensity is heavy-tailed (median 11 edits; max 1\,864).  These numbers corroborate the Pareto‐like intuition that a tiny fraction of lines soaks up a disproportionate maintenance budget.

Location matters. More than half of the hotspots live in administrative
files. Patterns such as \emph{Pinned Version Bump} and
\emph{Metadata Change} are almost exclusively administrative, whereas
\emph{Function Call Change} skews towards executable code.  Countermeasures
must therefore differ: configuration-management tooling for administrative
hotspots versus refactoring and architectural hardening for code hotspots.

\subsection{How does bot activity influence hotspots? (RQ3)}

Bots account for 73.9\% of the 20\,690 hotspot-related commits, originating from just 23 distinct automated accounts (Table~\ref{tab:bots}).  Their influence is pattern-specific: \emph{Metadata Change} is 95\% bot-generated, while five patterns
(\eg\ \emph{Service Configuration}, \emph{Stepwise Refactoring}) are entirely human.  Bots therefore concentrate on mechanical, high-frequency edits (semantic-version bumps, formatting tweaks), yet
humans remain dominant in semantically rich or context-dependent changes.  Filtering bot commits in dashboards and code-review tools would reduce noise and help teams focus on truly error-prone regions.

\subsection{Practice and Research Implications}
Practitioners can operationalize our taxonomy by translating recurring hotspot patterns into automated CI checks. A practical workflow is to periodically run our hotspot detector to identify dominant hotspot types, map these types to targeted CI rules, and enforce these rules on incoming pull requests to prevent avoidable churn.
For example, if developers observe repeated \emph{Formatting Ping-Pong} or \emph{Dependency Specification} changes, they could decide to refactor or establish more stable standards in the development process.
Teams can then track whether hotspot frequency decreases over time, providing quantitative feedback on mitigation effectiveness.

Future research could focus on empirically quantifying the impact of specific hotspot types on software quality metrics, such as defect rates and maintenance costs. Additionally, integrating machine learning techniques to automatically detect these hotspot types could provide developers with real-time insights, facilitating more informed decision-making in code management.
Furthermore, it is important to understand the role of automated bots as
hotspot generators.
As highlighted by \citet{erlenhov2020empirical}, bot-driven changes can sometimes introduce superficial modifications that inflate churn metrics, potentially masking significant development activities.

\section{Threats to Validity}
\label{sec:threats}
\paragraph{Internal Validity} Our line modification counting algorithm
can miss or misrepresent some changes.
During our manual labeling process we counted such changes---primarily
occuring in very large and complex \textit{diffs}.
Through this process we established that our algorithm
has an accuracy of 90\%, with 16 misreports out of 166.
Additionally, our tracker defines identity strictly by file path and line index. If developers move a block without editing its text, the original lines are logged as deleted and new ones as created, which shortens apparent lifespans and modification counts. Prior work based on tree-differencing can recover such moves~\cite{falleri2024fine}, but mainstream \texttt{git diff} does not expose that information at line granularity. 

\texttt{Git log} limitations may also affect the precision of our data extraction. 
Our operationalization of \emph{source code hotspots} as lines with modification counts exceeding three standard deviations above the mean, while aligned with established outlier detection approaches, represent a specific definition. Alternative thresholds or time-weighted modification frequencies might identify different sets of hotspots.

\paragraph{External Validity} In our study we mined repositories exclusively from GitHub, relying on GitHub's revision history. This dependency might limit the applicability of our findings to other version controls systems. Despite this limitation, our approach for identifying hotspots can still provide valuable insights about how developers make code changes across different development environments and platforms.

\section{Conclusion}
\label{sec:conclusion}
In this study, we have identified and categorized source code hotspots (line-level instances of code churn) in open-source software.
Through our analysis of 91 diverse GitHub repositories, we discovered 15 distinct hotspot types. Our findings provide two insights that challenge conventional assumptions about code churn. First, more than half of identified hotspots occur in administrative files rather than source code, indicating that configuration management practices significantly contribute to unnecessary revisions.
Second, bot activity accounts for 73.9\% of all analyzed commits, demonstrating the substantial impact of automated processes on code modifications.

Hotspots may indicate issues associated with software changeability, configurability, and stability---key quality attributes that directly influence software maintainability. By understanding these patterns, developers can proactively address issues that lead to frequent, avoidable changes in their codebases. To support this goal, we provide concrete mitigation strategies for each identified hotspot type, including semantic versioning, environment variables, linters, and cohesive refactoring planning.

By providing a tool to identify source code hotspots, categorizing their types, and documenting practical mitigation strategies, this research contributes to the broader understanding of software maintainability challenges while providing actionable guidance for creating more robust and stable software systems.

\paragraph{Data Availability}
To facilitate reproducibility, we provide: our dataset of labeled, hotspots (including identified bots); the complete source code for line-tracking, hotspot detection, and analysis; accompanying scripts for statistical computations; and detailed repository-selection criteria along with full repository metadata.
All these are available at \url{https://doi.org/10.5281/zenodo.15875147}.


\bibliographystyle{ACM-Reference-Format}
\bibliography{ref}

@article{buckley2005towards,
  author    = {Buckley, Jim and Mens, Tom and Zenger, Matthias and Rashid, Awais and Kniesel, G{\"u}nter},
  title     = {Towards a Taxonomy of Software Change},
  journal   = {Journal of Software Maintenance and Evolution: Research and Practice},
  volume    = {17},
  number    = {5},
  pages     = {309--332},
  year      = {2005},
  publisher = {Wiley},
  doi       = {10.1002/smr.319}
}

@inproceedings{mens2003towards,
  author    = {Mens, Tom and Buckley, Jim and Zenger, Matthias and Rashid, Awais},
  title     = {Towards a Taxonomy of Software Evolution},
  booktitle = {{IWUSE} '03: Proceedings of the International Workshop on Unanticipated Software Evolution},
  year      = {2003},
  publisher = {ACM}
}

@inproceedings{nguyen2013study,
  author    = {Nguyen, Hoan Anh and Nguyen, Anh Tuan and Nguyen, Tung Thanh and Nguyen, Tien N. and Rajan, Hridesh},
  title     = {A Study of Repetitiveness of Code Changes in Software Evolution},
  booktitle = {{ASE} '13: Proceedings of the 28th IEEE/ACM International Conference on Automated Software Engineering},
  pages     = {180--190},
  year      = {2013},
  publisher = {IEEE/ACM},
  doi       = {10.1109/ASE.2013.6693078}
}

@article{islam2021changes,
  author    = {Islam, Md Rakibul and Zibran, Minhaz F.},
  title     = {What Changes in Where? An Empirical Study of Bug-Fixing Change Patterns},
  journal   = {ACM SIGAPP Applied Computing Review},
  volume    = {20},
  number    = {4},
  pages     = {18--34},
  year      = {2021},
  publisher = {ACM},
  doi       = {10.1145/3447332.3447334}
}

@article{trautsch2023really,
  author    = {Trautsch, Alexander and Erbel, Johannes and Herbold, Steffen and Grabowski, Jens},
  title     = {What Really Changes When Developers Intend to Improve Their Source Code: A Commit-Level Study of Static Metric Value and Static Analysis Warning Changes},
  journal   = {Empirical Software Engineering},
  volume    = {28},
  number    = {2},
  pages     = {30},
  year      = {2023},
  publisher = {Springer},
  doi       = {10.1007/s10664-022-10257-9}
}

@inproceedings{fluri2008discovering,
  author    = {Fluri, Beat and Giger, Emanuel and Gall, Harald C.},
  title     = {Discovering Patterns of Change Types},
  booktitle = {{ASE} '08: Proceedings of the 23rd IEEE/ACM International Conference on Automated Software Engineering},
  pages     = {463--466},
  year      = {2008},
  publisher = {IEEE/ACM},
  doi       = {10.1109/ASE.2008.74}
}

@inproceedings{palomba2017exploratory,
  author    = {Palomba, Fabio and Zaidman, Andy and Oliveto, Rocco and De Lucia, Andrea},
  title     = {An Exploratory Study on the Relationship Between Changes and Refactoring},
  booktitle = {{ICPC} '17: Proceedings of the 25th IEEE/ACM International Conference on Program Comprehension},
  pages     = {176--185},
  year      = {2017},
  publisher = {IEEE/ACM},
  doi       = {10.1109/ICPC.2017.38}
}

@inproceedings{lehnert2012taxonomy,
  author    = {Lehnert, Steffen and Riebisch, Matthias},
  title     = {A Taxonomy of Change Types and Its Application in Software Evolution},
  booktitle = {{ECBS} '12: Proceedings of the 19th IEEE International Conference and Workshops on Engineering of Computer-Based Systems},
  pages     = {98--107},
  year      = {2012},
  publisher = {IEEE},
  doi       = {10.1109/ECBS.2012.9}
}

@inproceedings{mockus2000identifying,
  author    = {Mockus, Audris and Votta, Lawrence},
  title     = {Identifying Reasons for Software Changes Using Historic Databases},
  booktitle = {{ICSM} '00: Proceedings of the International Conference on Software Maintenance},
  pages     = {120--130},
  year      = {2000},
  publisher = {IEEE},
  doi       = {10.1109/ICSM.2000.883028}
}

@article{fluri2007change,
  author    = {Fluri, Beat and W{\"u}rsch, Michael and Pinzger, Martin and Gall, Harald},
  title     = {Change Distilling: Tree Differencing for Fine-Grained Source Code Change Extraction},
  journal   = {IEEE Transactions on Software Engineering},
  volume    = {33},
  number    = {11},
  pages     = {725--743},
  year      = {2007},
  publisher = {IEEE},
  doi       = {10.1109/TSE.2007.70731}
}

@inproceedings{lin2016empirical,
  author    = {Lin, Wei and Chen, Zhifei and Ma, Wanwangying and Chen, Lin and Xu, Lei and Xu, Baowen},
  title     = {An Empirical Study on the Characteristics of Python Fine-Grained Source-Code Change Types},
  booktitle = {{ICSME} '16: Proceedings of the IEEE International Conference on Software Maintenance and Evolution},
  pages     = {188--199},
  year      = {2016},
  publisher = {IEEE},
  doi       = {10.1109/ICSME.2016.25}
}

@article{jaafar2017analyzing,
  author    = {Jaafar, Fehmi and Lozano, Angela and Gu{\'e}h{\'e}neuc, Yann-Ga{\"e}l and Mens, Kim},
  title     = {Analyzing Software Evolution and Quality by Extracting Asynchrony Change Patterns},
  journal   = {Journal of Systems and Software},
  volume    = {131},
  pages     = {311--322},
  year      = {2017},
  publisher = {Elsevier},
  doi       = {10.1016/j.jss.2017.05.047}
}

@Book{Spi06,
  author    = {Spinellis, Diomidis},
  title     = {Code Quality: The Open Source Perspective},
  publisher = {Addison-Wesley},
  address   = {Boston, MA},
  year      = {2006},
  url       = {https://www.spinellis.gr/codequality},
  isbn      = {0-321-16607-8}
}

@book{brown1998,
  author    = {Brown, William H. and Malveau, Raphael C. and McCormick, Hays W. and Mowbray, Thomas J.},
  title     = {AntiPatterns: Refactoring Software, Architectures, and Projects in Crisis},
  publisher = {John Wiley \& Sons},
  year      = {1998}
}

@incollection{koenig1998,
  author    = {Koenig, Andrew},
  title     = {Patterns and AntiPatterns},
  booktitle = {The Patterns Handbooks: Techniques, Strategies, and Applications},
  pages     = {383--389},
  year      = {1998},
  publisher = {Wiley}
}

@book{webster1995,
  author    = {Webster, Bruce F.},
  title     = {Pitfalls of Object-Oriented Development},
  publisher = {M\&T Books},
  year      = {1995}
}

@book{fowler2018,
  author    = {Fowler, Martin},
  title     = {Refactoring: Improving the Design of Existing Code},
  publisher = {Addison-Wesley Professional},
  year      = {2018}
}

@article{mo2019,
  author    = {Mo, Ran and Cai, Yuanfang and Kazman, Rick and Xiao, Lu and Feng, Qiong},
  title     = {Architecture AntiPatterns: Automatically Detectable Violations of Design Principles},
  journal   = {IEEE Transactions on Software Engineering},
  volume    = {47},
  number    = {5},
  pages     = {1008--1028},
  year      = {2019},
  publisher = {IEEE},
  doi       = {10.1109/TSE.2019.2910856}
}

@inproceedings{garcia2009,
  author    = {Garcia, Joshua and Popescu, Daniel and Edwards, George and Medvidovic, Nenad},
  title     = {Identifying Architectural Bad Smells},
  booktitle = {{CSMR} '09: Proceedings of the 13th European Conference on Software Maintenance and Reengineering},
  pages     = {255--258},
  year      = {2009},
  publisher = {IEEE},
  doi       = {10.1109/CSMR.2009.59}
}

@article{taibi2018,
  author    = {Taibi, Davide and Lenarduzzi, Valentina},
  title     = {On the Definition of Microservice Bad Smells},
  journal   = {IEEE Software},
  volume    = {35},
  number    = {3},
  pages     = {56--62},
  year      = {2018},
  publisher = {IEEE},
  doi       = {10.1109/MS.2018.2141031}
}

@inproceedings{negara2014,
  author    = {Negara, Stas and Codoban, Mihai and Dig, Danny and Johnson, Ralph E.},
  title     = {Mining Fine-Grained Code Changes to Detect Unknown Change Patterns},
  booktitle = {{ICSE} '14: Proceedings of the 36th International Conference on Software Engineering},
  pages     = {803--813},
  year      = {2014},
  publisher = {ACM},
  doi       = {10.1145/2568225.2568317}
}

@inproceedings{palomba2013,
  author    = {Palomba, Fabio and Bavota, Gabriele and Di Penta, Massimiliano and Oliveto, Rocco and De Lucia, Andrea and Poshyvanyk, Denys},
  title     = {Detecting Bad Smells in Source Code Using Change-History Information},
  booktitle = {{ASE} '13: Proceedings of the 28th IEEE/ACM International Conference on Automated Software Engineering},
  pages     = {268--278},
  year      = {2013},
  publisher = {IEEE/ACM},
  doi       = {10.1109/ASE.2013.6693086}
}

@inproceedings{cai2019dv8,
  author    = {Cai, Yuanfang and Kazman, Rick},
  title     = {DV8: Automated Architecture Analysis Tool Suites},
  booktitle = {{TechDebt} '19: Proceedings of the IEEE/ACM International Conference on Technical Debt},
  pages     = {53--54},
  year      = {2019},
  publisher = {IEEE/ACM},
  doi       = {10.1109/TechDebt.2019.00015}
}

@inproceedings{sharma2016designite,
  author    = {Sharma, Tushar and Mishra, Pratibha and Tiwari, Rohit},
  title     = {Designite: A Software Design Quality Assessment Tool},
  booktitle = {{BADTDDA} '16: Proceedings of the 1st International Workshop on Bringing Architectural Design Thinking into Developers' Daily Activities},
  pages     = {1--4},
  year      = {2016},
  publisher = {ACM},
  doi       = {10.1145/2896935.2896938}
}

@Book{Hum89,
  author    = {Humphrey, Watts S.},
  title     = {Managing the Software Process},
  publisher = {Addison-Wesley},
  address   = {Reading, MA},
  year      = {1989},
  isbn      = {0201180952}
}

@Book{ISO18670:2025,
  author    = {{International Organization for Standardization}},
  title     = {Information Technology — SoftWare Hash IDentifier ({SWHID}) Specification V1.2},
  note      = {ISO/IEC 18670:2025},
  address   = {Geneva, Switzerland},
  year      = {2025}
}

@inproceedings{sharma2016does,
  author    = {Sharma, Tushar and Fragkoulis, Marios and Spinellis, Diomidis},
  title     = {Does Your Configuration Code Smell?},
  booktitle = {{MSR} '16: Proceedings of the 13th IEEE/ACM Working Conference on Mining Software Repositories},
  pages     = {189--200},
  year      = {2016},
  publisher = {IEEE/ACM},
  doi       = {10.1145/2901739.2901761}
}

@article{palomba2014mining,
  author    = {Palomba, Fabio and Bavota, Gabriele and Di Penta, Massimiliano and Oliveto, Rocco and Poshyvanyk, Denys and De Lucia, Andrea},
  title     = {Mining Version Histories for Detecting Code Smells},
  journal   = {IEEE Transactions on Software Engineering},
  volume    = {41},
  number    = {5},
  pages     = {462--489},
  year      = {2014},
  publisher = {IEEE},
  doi       = {10.1109/TSE.2014.2372760}
}

@article{miller1985file,
  author    = {Miller, W. Eugene and Myers, W.},
  title     = {A File Comparison Program},
  journal   = {Software: Practice and Experience},
  volume    = {15},
  number    = {11},
  pages     = {1040--1044},
  year      = {1985},
  publisher = {Wiley}
}

@inproceedings{asaduzzaman2013lhdiff,
  author    = {Asaduzzaman, Muhammad and Roy, Chanchal K. and Schneider, Kevin A. and Di Penta, Massimiliano},
  title     = {LHDiff: A Language-Independent Hybrid Approach for Tracking Source Code Lines},
  booktitle = {{ICSM} '13: Proceedings of the 29th IEEE International Conference on Software Maintenance},
  pages     = {230--239},
  year      = {2013},
  publisher = {IEEE}
}

@inproceedings{falleri2014fine,
  author    = {Falleri, Jean-R{\'e}my and Morandat, Flor{\'e}al and Blanc, Xavier and Martinez, Matias and Monperrus, Martin},
  title     = {Fine-Grained and Accurate Source-Code Differencing},
  booktitle = {{ASE} '14: Proceedings of the 29th ACM/IEEE International Conference on Automated Software Engineering},
  pages     = {313--324},
  year      = {2014},
  publisher = {ACM/IEEE}
}

@article{myers1986nd,
  author    = {Myers, Eugene W.},
  title     = {An {$O(ND)$} Difference Algorithm and Its Variations},
  journal   = {Algorithmica},
  volume    = {1},
  number    = {1},
  pages     = {251--266},
  year      = {1986},
  publisher = {Springer}
}

@inproceedings{falleri2024fine,
  author    = {Falleri, Jean-R{\'e}my and Martinez, Matias},
  title     = {Fine-Grained, Accurate and Scalable Source Differencing},
  booktitle = {{ICSE} '24: Proceedings of the 46th International Conference on Software Engineering},
  pages     = {1--12},
  year      = {2024},
  publisher = {IEEE/ACM}
}

@article{smith1984nonparametric,
  author    = {Smith, Eric P. and van Belle, Gerald},
  title     = {Nonparametric Estimation of Species Richness},
  journal   = {Biometrics},
  pages     = {119--129},
  year      = {1984},
  publisher = {Wiley}
}

@article{borges2018s,
  author    = {Borges, Hudson and Valente, Marco Tulio},
  title     = {What’s in a GitHub Star? Understanding Repository Starring Practices in a Social Coding Platform},
  journal   = {Journal of Systems and Software},
  volume    = {146},
  pages     = {112--129},
  year      = {2018},
  publisher = {Elsevier},
  doi       = {10.1016/j.jss.2018.09.016}
}

@inproceedings{hassan2009,
  author    = {Hassan, Ahmed E.},
  title     = {Predicting Faults Using the Complexity of Code Changes},
  booktitle = {{ICSE} '09: Proceedings of the 31st International Conference on Software Engineering},
  pages     = {78--88},
  year      = {2009},
  publisher = {IEEE},
  doi       = {10.1109/ICSE.2009.5070510}
}

@inproceedings{giger2012,
  author    = {Giger, Emanuel and Pinzger, Martin and Gall, Harald C.},
  title     = {Can We Predict Types of Code Changes? An Empirical Analysis},
  booktitle = {{MSR} '12: Proceedings of the 9th IEEE Working Conference on Mining Software Repositories},
  pages     = {217--226},
  year      = {2012},
  publisher = {IEEE}
}

@article{wessel2018,
  author    = {Wessel, Mairieli and De Souza, Bruno Mendes and Steinmacher, Igor and Wiese, Igor S. and Polato, Ivanilton and Chaves, Ana Paula and Gerosa, Marco A.},
  title     = {The Power of Bots: Characterising and Understanding Bots in OSS Projects},
  journal   = {Proceedings of the ACM on Human-Computer Interaction},
  volume    = {2},
  number    = {CSCW},
  pages     = {1--19},
  year      = {2018},
  publisher = {ACM},
  doi       = {10.1145/3274451}
}

@inproceedings{dey2020exploratory,
  author    = {Dey, Tapajit and Vasilescu, Bogdan and Mockus, Audris},
  title     = {An Exploratory Study of Bot Commits},
  booktitle = {{ICSE} '20: Proceedings of the 42nd International Conference on Software Engineering Workshops},
  pages     = {61--65},
  year      = {2020},
  publisher = {IEEE/ACM},
  doi       = {10.1145/3387940.3391502}
}

@article{vidoni2022systematic,
  author    = {Vidoni, Melina},
  title     = {A Systematic Process for Mining Software Repositories: Results from a Systematic Literature Review},
  journal   = {Information and Software Technology},
  volume    = {144},
  pages     = {106791},
  year      = {2022},
  publisher = {Elsevier},
  doi       = {10.1016/j.infsof.2021.106791}
}

@inproceedings{dabic2021sampling,
  author    = {Dabic, Ozren and Aghajani, Emad and Bavota, Gabriele},
  title     = {Sampling Projects in GitHub for MSR Studies},
  booktitle = {{MSR} '21: Proceedings of the 18th IEEE/ACM International Conference on Mining Software Repositories},
  pages     = {560--564},
  year      = {2021},
  publisher = {IEEE/ACM},
  doi       = {10.1109/MSR52588.2021.00074}
}

@inproceedings{erlenhov2020empirical,
  author    = {Erlenhov, Linda and Gomes De Oliveira Neto, Francisco and Leitner, Philipp},
  title     = {An Empirical Study of Bots in Software Development: Characteristics and Challenges from a Practitioner’s Perspective},
  booktitle = {{ESEC/FSE} '20: Proceedings of the 28th ACM Joint Meeting on European Software Engineering Conference and Symposium on the Foundations of Software Engineering},
  pages     = {445--455},
  year      = {2020},
  publisher = {ACM},
  doi       = {10.1145/3368089.3409680}
}

@article{dohmke2023100,
  author    = {Dohmke, Thomas},
  title     = {100 Million Developers and Counting},
  journal   = {The GitHub Blog},
  volume    = {25},
  year      = {2023}
}

@inproceedings{kinsman2021software,
  author    = {Kinsman, Timothy and Wessel, Mairieli and Gerosa, Marco A. and Treude, Christoph},
  title     = {How Do Software Developers Use GitHub Actions to Automate Their Workflows?},
  booktitle = {{MSR} '21: Proceedings of the 18th IEEE/ACM International Conference on Mining Software Repositories},
  pages     = {420--431},
  year      = {2021},
  publisher = {IEEE/ACM},
  doi       = {10.1109/MSR52588.2021.00054}
}

@article{golzadeh2021ground,
  author    = {Golzadeh, Mehdi and Decan, Alexandre and Legay, Damien and Mens, Tom},
  title     = {A Ground-Truth Dataset and Classification Model for Detecting Bots in GitHub Issue and PR Comments},
  journal   = {Journal of Systems and Software},
  volume    = {175},
  pages     = {110911},
  year      = {2021},
  publisher = {Elsevier},
  doi       = {10.1016/j.jss.2021.110911}
}

@article{landis1977measurement,
  author    = {Landis, J. Richard and Koch, Gary G.},
  title     = {The Measurement of Observer Agreement for Categorical Data},
  journal   = {Biometrics},
  pages     = {159--174},
  year      = {1977},
  publisher = {Wiley},
  doi       = {10.2307/2529310}
}

@article{munaiah2017curating,
  author    = {Munaiah, Nuthan and Kroh, Steven and Cabrey, Craig and Nagappan, Meiyappan},
  title     = {Curating GitHub for Engineered Software Projects},
  journal   = {Empirical Software Engineering},
  volume    = {22},
  pages     = {3219--3253},
  year      = {2017},
  publisher = {Springer},
  doi       = {10.7287/peerj.preprints.2617v1}
}

@article{graves2002predicting,
  author    = {Graves, Todd L. and Karr, Alan F. and Marron, James S. and Siy, Harvey},
  title     = {Predicting Fault Incidence Using Software Change History},
  journal   = {IEEE Transactions on Software Engineering},
  volume    = {26},
  number    = {7},
  pages     = {653--661},
  year      = {2002},
  publisher = {IEEE},
  doi       = {10.1109/32.859533}
}

@inproceedings{nagappan2005use,
  author    = {Nagappan, Nachiappan and Ball, Thomas},
  title     = {Use of Relative Code-Churn Measures to Predict System Defect Density},
  booktitle = {{ICSE} '05: Proceedings of the 27th International Conference on Software Engineering},
  pages     = {284--292},
  year      = {2005},
  publisher = {IEEE},
  doi       = {10.1109/ICSE.2005.1553571}
}

@inproceedings{kim2007predicting,
  author    = {Kim, Sunghun and Zimmermann, Thomas and Whitehead Jr., E. James and Zeller, Andreas},
  title     = {Predicting Faults from Cached History},
  booktitle = {{ICSE} '07: Proceedings of the 29th International Conference on Software Engineering},
  pages     = {489--498},
  year      = {2007},
  publisher = {IEEE},
  doi       = {10.1145/1342211.1342216}
}

@inproceedings{kalliamvakou2014promises,
  author    = {Kalliamvakou, Eirini and Gousios, Georgios and Blincoe, Kelly and Singer, Leif and German, Daniel M. and Damian, Daniela},
  title     = {The Promises and Perils of Mining GitHub},
  booktitle = {{MSR} '14: Proceedings of the 11th Working Conference on Mining Software Repositories},
  pages     = {92--101},
  year      = {2014},
  publisher = {ACM},
  doi       = {10.1145/2597073.2597074}
}

@article{brereton2007lessons,
  title={Lessons from applying the systematic literature review process within the software engineering domain},
  author={Brereton, Pearl and Kitchenham, Barbara A and Budgen, David and Turner, Mark and Khalil, Mohamed},
  journal={Journal of systems and software},
  volume={80},
  number={4},
  pages={571--583},
  year={2007},
  publisher={Elsevier},
  doi={10.1016/j.jss.2006.07.009}
}

@article{HMT13,
	title={Application of Infrared Thermography for Predictive/Preventive Maintenance of Thermal Defect in Electrical Equipment},
	volume={61},
	ISSN={1359-4311},
	DOI={10.1016/j.applthermaleng.2013.07.028},
	number={2},
	journal={Applied Thermal Engineering},
	publisher={Elsevier BV},
	author={Huda, A.S. Nazmul and Taib, Soib},
	year={2013},
	month=nov,
	pages={220–227},
}

@article{JST12,
	title={Recent Progress in Diagnosing the Reliability of Electrical Equipment by Using Infrared Thermography},
	volume={55},
	ISSN={1350-4495},
	DOI={10.1016/j.infrared.2012.03.002},
	number={4},
	journal={Infrared Physics \& Technology},
	publisher={Elsevier BV},
	author={Jadin, Mohd Shawal and Taib, Soib},
	year={2012},
	month=jul,
	pages={236–245},
}

@article{APWM21,
	title={Raincloud plots: a multi-platform tool for robust data visualization},
	volume={4},
	ISSN={2398-502X},
	DOI={10.12688/wellcomeopenres.15191.2},
	journal={Wellcome Open Research},
	publisher={F1000 Research Ltd},
	author={Allen, Micah and Poggiali, Davide and Whitaker, Kirstie and Marshall, Tom Rhys and van Langen, Jordy and Kievit, Rogier A.},
	year={2021},
	month=jan,
	pages={63},
}

@TechReport{Ker79,
        Author="Brian W. Kernighan",
        Title="{UNIX} for Beginners",
        Number=75,
        Type="Computer Science Technical Report",
        URL="https://web.archive.org/web/20170711222622/http://wolfram.schneider.org/bsd/7thEdManVol2/beginners/beginners.pdf",
        Note="Available online \url{https://web.archive.org/web/20170711222622/http://wolfram.schneider.org/bsd/7thEdManVol2/beginners/beginners.pdf}",
        Month=feb,
        Year=1979,
        Institution="Bell Laboratories",
        Address="Murray Hill, NJ",
        Where="/pub/lib"
}

@article{MCKX21,
	title={Architecture Anti-Patterns: Automatically Detectable Violations of Design Principles},
	volume={47},
	ISSN={2326-3881},
	DOI={10.1109/tse.2019.2910856},
	number={5},
	journal={IEEE Transactions on Software Engineering},
	publisher={Institute of Electrical and Electronics Engineers (IEEE)},
	author={Mo, Ran and Cai, Yuanfang and Kazman, Rick and Xiao, Lu and Feng, Qiong},
	year={2021},
	month=may,
	pages={1008–1028},
}

@INPROCEEDINGS{JAJ12,
  author={Jamilah Din and AL-Badareen, Anas Bassam and Jusoh, Yusmadi Yah},
  booktitle={2012 7th International Conference on Computing and Convergence Technology (ICCCT)}, 
  title={Antipatterns detection approaches in Object-Oriented Design: A literature review}, 
  year={2012},
  pages={926-931},
}

@incollection{PDBO14,
	title={Anti-Pattern Detection},
	ISSN={0065-2458},
	DOI={10.1016/b978-0-12-800160-8.00004-8},
	booktitle={Advances in Computers},
	publisher={Elsevier},
	author={Palomba, Fabio and De Lucia, Andrea and Bavota, Gabriele and Oliveto, Rocco},
	year={2014},
	pages={201–238},
}

@inproceedings{PHBE22,
	series={EuroPLop ’22},
	title={Process anti-pattern detection – a case study},
	DOI={10.1145/3551902.3551965},
	booktitle={Proceedings of the 27th European Conference on Pattern Languages of Programs},
	publisher={ACM},
	author={Pícha, Petr and Hönel, Sebastian and Brada, Přemek and Ericsson, Morgan and Löwe, Welf and Wingkvist, Anna and Daněk, Jakub},
	year={2022},
	month=jul,
	pages={1–18},
	collection={EuroPLop ’22},
}

@inbook{Ney92,
	title={On the Two Different Aspects of the Representative Method: the Method of Stratified Sampling and the Method of Purposive Selection},
	ISBN={9781461243809},
	ISSN={0172-7397},
	DOI={10.1007/978-1-4612-4380-9_12},
	booktitle={Breakthroughs in Statistics},
	publisher={Springer New York},
	author={Neyman, Jerzy},
	year={1992},
	pages={123–150},
}

@InProceedings{SLMS24,
	author="Diomidis Spinellis and Panos Louridas and Maria Kechagia and Tushar Sharma",
	title="Broken Windows: {E}xploring the Applicability of a Controversial Theory on Code Quality",
	booktitle="ICSME '24: International Conference on Software Maintenance and Evolution",
	location="Flagstaff, AZ, USA",
	month=oct,
	pages="400--412",
	URL="https://doi.org/10.48550/arXiv.2410.13480",
	doi="10.1109/ICSME58944.2024.00044",
	year="2024",
	publisher="IEEE",
}

@article{SLK21,
	title = {Software Evolution: The Lifetime of Fine-Grained Elements},
	doi = {10.7717/peerj-cs.372},
	year = {2021},
	month = feb,
	publisher = {PeerJ},
	volume = {7},
	pages = {e372},
	author = {Diomidis Spinellis and Panos Louridas and Maria Kechagia},
	journal = {{PeerJ} Computer Science},
	tags="sweng,bdp",
	oa = 1,
}

@book{FGK15,
	title={Academic Work and Careers in {E}urope: Trends, Challenges, Perspectives},
	Editor={Tatiana Fumasoli and Ga\"{e}le Goastellec and Barbara M. Kehm},
	ISBN={9783319107202},
	DOI={10.1007/978-3-319-10720-2},
	publisher={Springer International Publishing},
	year={2015},
}

@inproceedings{DZ17,
  title={Software heritage: Why and how to preserve software source code},
  author={Di Cosmo, Roberto and Zacchiroli, Stefano},
  booktitle={iPRES 2017-14th International Conference on Digital Preservation},
  pages={1--10},
  year={2017}
}

@article{All11,
	title={The robustness principle reconsidered},
	volume={54},
	ISSN={1557-7317},
	DOI={10.1145/1978542.1978557},
	number={8},
	journal={Communications of the ACM},
	publisher={Association for Computing Machinery (ACM)},
	author={Allman, Eric},
	year={2011},
	month=aug,
	pages={40–45},
}

@Book{GHJV95,
	Title="Design Patterns: Elements of Reusable Object-Oriented Software",
	Author="Erich Gamma and Richard Helm and Ralph Johnson and John Vlissides",
	Publisher="Addison-Wesley",
	Address="Reading, MA",
	Year="1995",
	DDC = "005.12",
	ISBN="0-201-63361-2"
	}

@article{LZ22,
	title={Reproducible Builds: Increasing the Integrity of Software Supply Chains},
	volume={39},
	ISSN={1937-4194},
	DOI={10.1109/ms.2021.3073045},
	number={2},
	journal={IEEE Software},
	publisher={Institute of Electrical and Electronics Engineers (IEEE)},
	author={Lamb, Chris and Zacchiroli, Stefano},
	year={2022},
	month=mar,
	pages={62–70},
}

@inproceedings{KRS13,
        series={WikiSym ’13},
        title={The empirical commit frequency distribution of open source
projects},
        DOI={10.1145/2491055.2491073},
        booktitle={Proceedings of the 9th International Symposium on Open
Collaboration},
        publisher={ACM},
        author={Kolassa, Carsten and Riehle, Dirk and Salim, Michel A.},
        year={2013},
        month=aug,
        pages={1–8},
        collection={WikiSym ’13},
}

@inproceedings{XWWZ22,
        title={Exploring Activity and Contributors on {GitHub}: Who, What,
When, and Where},
        DOI={10.1109/apsec57359.2022.00013},
        booktitle={2022 29th Asia-Pacific Software Engineering Conference (APSEC)},
        publisher={IEEE},
        author={Xia, Xiaoya and Weng, Zhenjie and Wang, Wei and Zhao, Shengyu},
        year={2022},
        month=dec,
        pages={11–20},
}

@article{miller1999estimating,
  title={Estimating the number of remaining defects after inspection},
  author={Miller, James},
  journal={Software Testing, Verification and Reliability},
  volume={9},
  number={3},
  pages={167--189},
  year={1999},
  publisher={Wiley Online Library}
}

@article{bohme2018stads,
  title={{STADS}: Software testing as species discovery},
  author={B{\"o}hme, Marcel},
  journal={ACM Transactions on Software Engineering and Methodology},
  volume={27},
  number={2},
  pages={1--52},
  year={2018},
  publisher={ACM New York, NY, USA}
}

@inproceedings{liyanage2023reachable,
  title={Reachable coverage: Estimating saturation in fuzzing},
  author={Liyanage, Danushka and B{\"o}hme, Marcel and Tantithamthavorn, Chakkrit and Lipp, Stephan},
  booktitle={ICSE '23: 45th IEEE/ACM International Conference on Software Engineering},
  pages={371--383},
  year={2023},
  organization={IEEE}
}

@inproceedings{MCKX15,
	title={Hotspot Patterns: The Formal Definition and Automatic Detection of Architecture Smells},
	DOI={10.1109/wicsa.2015.12},
	booktitle={2015 12th Working IEEE/IFIP Conference on Software Architecture},
	publisher={IEEE},
	author={Mo, Ran and Cai, Yuanfang and Kazman, Rick and Xiao, Lu},
	year={2015},
	month=may,
	pages={51–60},
}

@article{SAKS20,
	title={Exploring the Relation Between Co-changes and Architectural Smells},
	volume={2},
	ISSN={2661-8907},
	DOI={10.1007/s42979-020-00407-5},
	number={1},
	journal={SN Computer Science},
	publisher={Springer Science and Business Media LLC},
	author={Sas, Darius and Avgeriou, Paris and Kruizinga, Ronald and Scheedler, Ruben},
	year={2020},
	month=dec
}

\end{document}